\documentclass[fleqn,usenatbib]{mnras}

\defcitealias{2022lopeznavas}{LN22}
\usepackage{newtxtext,newtxmath}

\usepackage[T1]{fontenc}
\usepackage{subcaption}
\usepackage{adjustbox}
\usepackage{threeparttable}
\usepackage{comment}
\usepackage{ulem}

\DeclareRobustCommand{\VAN}[3]{#2}
\let\VANthebibliography\thebibliography
\def\thebibliography{\DeclareRobustCommand{\VAN}[3]{##3}\VANthebibliography}

\usepackage{graphicx}	
\usepackage{amsmath}	
\usepackage{multicol}

\title[Multi-wavelength variability in changing-looks]{Improving the selection of changing-look AGNs through multi-wavelength photometric variability}
\author[E. López-Navas]{E. López-Navas$^{1,2}$\thanks{E-mail: elena.lopez@postgrado.uv.cl},
P. Sánchez-Sáez$^{3,4}$,
P. Arévalo$^{1,2}$,
S. Bernal$^{1}$,
M. J. Graham$^{5}$,
\newauthor
L. Hernández-García$^{4,1}$,
D. Homan$^{6}$,
M. Krumpe$^{6}$,
G. Lamer$^{6}$,
P. Lira$^{7}$,
M.L. Martínez-Aldama$^{1,2}$,
\newauthor
A. Merloni$^{8}$,
S. Ríos$^{7}$,
M. Salvato$^{8,9}$,
D. Stern$^{10}$ and 
D. Tubín-Arenas$^{6}$.
\\
$^{1}$Instituto de Física y Astronomía, Facultad de Ciencias, Universidad de Valparaíso, Gran Bretaña 1111, Valparaíso, Chile\\
$^{2}$Millennium Nucleus on Transversal Research and Technology to Explore Supermassive Black Holes (TITANS)\\
$^{3}$European Southern Observatory, Karl-Schwarzschild-Strasse 2, 85748 Garching bei München, Germany\\
$^{4}$Millennium Institute of Astrophysics (MAS), Nuncio Monseñor Sótero Sanz 100, Providencia, Santiago, Chile\\
$^{5}$California Institute of Technology, 1200 E. California Blvd, Pasadena, CA 91125, USA\\
$^{6}$Leibniz-Institut für Astrophysik Potsdam (AIP), An der Sternwarte 16, 14482 Potsdam, Germany \\
$^{7}$Departamento Astronomía, Universidad de Chile, Casilla 36D, Santiago, Chile\\
$^{8}$Max-Planck-Institut für extraterrestrische Physik, Giessenbachstr. 1, 85748 Garching, Germany\\
$^{9}$Exzellenzcluster ORIGINS, Boltzmannstr. 2, 85748 Garching, Germany \\
$^{10}$Jet Propulsion Laboratory, California Institute of Technology, 4800 Oak Grove Drive, Pasadena, CA 91109, USA
}

\date{Accepted XXX. Received YYY; in original form ZZZ}

\pubyear{2023}

\begin{document}
\label{firstpage}
\pagerange{\pageref{firstpage}--\pageref{lastpage}}
\maketitle

\begin{abstract}


We present second epoch optical spectra for 30 changing-look (CL) candidates found by searching for Type-1 optical variability in a sample of active galactic nuclei (AGNs) spectroscopically classified as Type 2. We use a random-forest-based light curve classifier and spectroscopic follow-up, confirming 50 per cent of candidates as turning-on CLs. In order to improve this selection method and to better understand the nature of the not-confirmed CL candidates, we perform a multi-wavelength variability analysis including optical, mid-infrared (MIR) and X-ray data, and compare the results from the confirmed and not-confirmed CLs identified in this work. We find that most of the not-confirmed CLs are consistent with weak Type 1s dominated by host-galaxy contributions, showing weaker optical and MIR variability. On the contrary, the confirmed CLs present stronger optical fluctuations and experience a long (from five to ten years) increase in their MIR fluxes and the colour \textit{W1--W2} over time. In the 0.2--2.3 keV band, at least four out of 11 CLs with available \textit{SRG}/eROSITA detections have increased their flux in comparison with archival upper limits. These common features allow us to select the most promising CLs from our list of candidates, leading to nine sources with similar multi-wavelength photometric properties to our CL sample. The use of machine learning algorithms with optical and MIR light curves will be very useful to identify CLs in future large-scale surveys.



\end{abstract}

\begin{keywords}
galaxies: active -- accretion, accretion discs -- quasars: emission lines
\end{keywords}



\section{Introduction}

For the past few years, a growing (>200) population of active galactic nuclei (AGNs) with emerging or disappearing optical broad emission lines (BELs) has been found, arousing great interest among the astrophysics community \citep[see review by ][]{2022review}. 
Most studies favour an accretion rate change as the origin of such dramatic changes in unobscured AGNs, so these sources are often called changing-state (CS) AGNs. Other mechanisms such as variable absorption and tidal disruption events (TDEs) are also expected to produce variations in the BELs, so the term changing-look (CL) is generally used to refer to all AGNs that show such spectral transitions, regardless of the physical mechanisms driving these changes. This term is borrowed from the X-ray community, where a CL event is led by extreme variable X-ray absorption, causing a switch between Compton-thin ($N_{\rm H}$ < 10$^{24}$ cm$^{-2}$) and Compton-thick ($N_{\rm H}$ $\gtrsim$ 10$^{24}$ cm$^{-2}$) states in AGNs \citep[e.g.][]{2003matt}.  


The CL phenomenon is characterised by drastic changes in the optical BELs. The BELs consist of permitted and semi-forbidden emission lines with typical line widths FWHM $\geq$ 1000 km s$^{-1}$, formed by high density gas clouds called the broad line region (BLR) located close to the central engine \citep[e.g. ][]{2015netzer}. Therefore, most of the effort to find CL AGNs has focused on systematic searches of broad Balmer line variations (generally >3$\sigma$ flux change in broad H~$\beta$) in sources with multi-epoch spectroscopy \citep[although other lines such as Mg \textsc{ii} are also possible, see ][]{macleod2016, ross2018, 2019guo}. In particular, some sizable samples have been found comparing repeated spectra from different surveys such as the Sloan Digital Sky Survey (SDSS) and the Large Sky Area Multi-Object Fiber Spectroscopic Telescope (LAMOST) \citep{yang2018, 2022green}. 

CL events are often accompanied by large photometric changes in the optical and ultraviolet (UV) bands, and this has been used as a selection criteria to find new CL AGNs \citep[i.e., |$\Delta$\textit{g}| > 1 mag, |$\Delta$\textit{r}| > 0.5 mag in ][]{macleod2016,macleod2019}. However, the link between extreme spectroscopic and photometric changes is uncertain, since just 10--15 per cent of photometrically variable AGNs have been found to display CL behaviour \citep{macleod2019}. This uncertainty can be affected by the time-scales involved in CL events, which have been constrained just in a few sources \citep{trakhtenbrot2019}, and the fact that the CL behaviour has been found to occur repeatedly in some sources \citep[e.g. depending on the Eddington ratio, ][]{2021guolo}.   

More recently, some studies have concentrated on the search for CL events based on the physical expectations for an accretion-state change. In this scenario, it is expected to see gradual changes in the optical and mid-infrared (MIR) flux and colours associated with monotonically varying BEL strengths and/or continuum changes, as the AGN goes to bright (AGN dominated) or dim (host dominated) states \citep[e.g.,][]{2017sheng, yang2018, 2022lyu}. In the optical, CS AGN candidates have been selected by searching for anomalous variability \citep{sanchez2021_anomaly} and bluer optical colours in turning-on AGNs \citep{2022hon}, where the latter method shows a higher success rate for confirmed CL compared to other selection techniques. In the MIR band, individual CL AGNs have been found by identifying highly MIR-variable quasars in the \textit{Wide-field Infrared Survey Explorer} (\textit{WISE}) and Near-Earth Object WISE Reactivation (NEOWISE) data stream  \citep{2018stern,2018assef}.
In \citet{graham2020}, 111 CS quasars were found by applying two different criteria: strongly enhanced optical variability  over some time-scales and a large absolute difference between  the  initial  and  final  state in the \textit{WISE} light curve  (i.e, $|\Delta  \textit{W1} |>0.2$ or $|\Delta  \textit{W2} |>0.2 $). That work led to a CS sample at higher luminosity than previous CL AGNs in the literature.


Moreover, individual CL events have been associated with changes in soft-X-ray/\textit{UV} emission, responsible for photoionizing the BLR gas, as in the case of Mrk 1018 \citep{1986cohen, 2016mcelroy, 2018noda}. More extreme X-ray spectral and flux variability was found in the CL source 1ES 1927+654, which has been suggested to be caused by a TDE in an AGN \citep{trakhtenbrot2019, ricci2020, 2021ricci} or a magnetic flux inversion event \citep{2022laha}. Recently, in a CL search of sources with multi-epoch optical spectroscopy within the \textit{Swift}-BAT AGN Spectroscopic Survey (BASS), it was reported that five out of nine events with \textit{Swift}-BAT data available could be associated with significant flux changes in the 14–195 keV hard X-ray band \citep{2022temple}.






With the advent of deep, large sky-coverage monitoring surveys such as the Zwicky Transient Facility \citep[ZTF,][]{bellm2019} and the upcoming Legacy Survey of Space and Time \citep[LSST,][]{2019lsst}, the identification of CL AGNs will be possible using machine-learning techniques. In \citet[][hereafter LN22]{2022lopeznavas}, we present a method specifically looking for turn-on events using a balanced random forest algorithm with the ZTF alert stream \citep{sanchez2021}, confirming CL behavior in four out of six sources that we re-observed based on follow-up spectroscopy. Extending this work further, we obtained a second epoch spectra of 30 additional CL candidates, confirming $\sim$50 per cent as CLs. In this paper, we present the new observations and perform a multi-wavelength (optical, MIR and X-ray) variability analysis of the CL sources. This effort enables us to improve the selection technique and reinforces the common features of these CL events. Throughout this work, we assume a standard cosmological model with $H_0$ = 70 km s$^{-1}$ Mpc$^{-1}$, $\Omega_{m}$ = 0.3, and $\Omega_{\Lambda}$ = 0.7.



\section{Selection of the sample}\label{sec:sample} 
According to the classic view of AGNs, the optical variability in Type 2 sources is highly suppressed due to obscuration of the continuum coming from the central source by the dusty torus \citep{antonucci1993}. Based on this consideration, looking for Type 1-like optical flux variability (coming from the accretion disc) in spectrally-classified Type 2 AGNs (whose accretion disc \textit{should} be hidden) has led to the finding of turning-on CL sources. We note that in these cases, the previous Type 2 classification is due to the absence of significant BELs in their spectra, and not due to the identification of the viewing angle of the system. This is the selection strategy we followed in \citetalias{2022lopeznavas} to find potential CL candidates. Here, we update the candidate list reported in \citetalias{2022lopeznavas} and clean it further.

Our initial sample consists of all the spectrally classified Type 2 AGNs included in the Million Quasars Catalog \citep[MILLIQUAS, Version 7.7, N and K types, ][]{milliquas}. We removed sources classified as Seyfert 1, Low Ionization Nuclear Emission Region (LINER) or blazar in any other study according to the SIMBAD Astronomical Database, and those included in the Type 1 AGN catalogues from \citet{Oh2015} and \citet{2019liu}. We also required the sources to have a public `GALAXY AGN' or `QSO AGN' spectrum in SDSS DR17, and discarded objects with a BROADLINE classification. This led to 20834 Type 2 AGNs. We also checked that none of these sources were included in the Roma-BZCAT Multifrequency Catalogue of Blazars \citep{2015massaro}. 

To select potential CL candidates, we searched for \textit{current} Type 1 optical flux variability in our Type 2 sample. In particular, we used the host- and core-dominated AGN classifications given by the Automatic Learning for the Rapid Classification of Events \citep[ALeRCE,][]{forster2021} broker light curve classifier \citep[LCC, ][]{sanchez2021}. The ALeRCE broker is currently processing the alert stream from the Zwicky Transient Facility \citep[ZTF,][]{bellm2014,bellm2019} to provide a fast classification of variable objects by applying a balanced random forest algorithm. In particular, the LCC computes a total of 174 features, including colours obtained from AllWISE and ZTF photometry and variability features, for all objects that had generated at least 6 alerts in either \textit{g} or \textit{r} band in the ZTF alert stream. Since each alert is produced when a 5$\sigma$ variation in the template-subtracted image occurs, only sufficiently variable objects are detected and classified by the LCC. In the case of AGNs, only 10 per cent of known Type 1s with $r<20.5$ mag exhibit variations reaching this threshold and produce alerts in ZTF. Therefore, using this method we expect to select the 10 per cent most variable CLs in the parent sample. 

We performed a sky crossmatch within 1 arcsec between our Type 2 parent sample and the sources classified primarily as AGN or QSO by the LCC (updated on 2022 November 30), which led to 71 matches. Ten sources had two different identification names in ZTF, resulting in 61 CL candidates. Of these, $\sim$30 objects were identified as bad candidates from a visual examination of the optical light curves and the SDSS spectra. In the first place, some sources show BELs in their optical spectra and were apparently misclassified in the catalogues. This generally occurs in the lower-luminosity and lower-black hole-mass regimes, where the BELs fail to meet the FWHM $\geq$ 1000 km s$^{-1}$ criterion \citep{2019liu}, and also in intermediate Type 1.8/1.9 AGNs, which show weak broad Balmer lines \citep[e.g.][]{2017hernandez}. Secondly, other sources appear misclassified by the ALeRCE LCC due to a small number of data points or transient events in the reference image used to construct the difference images \citep[mostly supernovae,][]{2023sanchez}. 


We note that the ZTF alert light curves only contain alerts generated since the ZTF started its operations in 2018, but the actual length of the alert light curves depend on how variable the object is. We stress that the first ZTF alert does not necessarily mark the time of the change of state, considering that most known Type 1 AGNs have not triggered alerts or have taken years after the start of ZTF to show their first alert. This method does not give information about time the CL transition occurred other than that it happened some time between the SDSS spectrum was taken 10--20 years ago and the ALeRCE LCC classification as Type 1 AGN. In order to confirm the selected candidates as CLs, we need to perform optical follow-up spectroscopy that quantifies the changes in the BELs with respect to the SDSS spectrum.

In this work, we further investigate the variability and spectral properties of the most promising CL candidates to improve the selection method and shed light on the origin of these phenomena.



\section{Spectroscopic follow-up}
\subsection{Data}

We obtained second epoch spectra for 36 of our 61 CL candidates, allowing us to confirm CL objects by quantifying BEL changes with respect to archival SDSS spectra. These sources were selected via visual examination of the optical light curves and  archival spectra of the CL candidates. Six out of the 36 sources were reported in \citetalias{2022lopeznavas}. In this paper, we present spectral analysis for the remaining 30 objects.  The new optical spectra were taken during February and April 2022 using either the Double Spectrograph (DBSP) on the Palomar 200-inch Hale Telescope (P200) or  the  Low  Resolution  Imaging Spectrometer (LRIS) spectrograph on the Keck I telescope at the W. M. Keck Observatory, as specified in the Appendix, Table \ref{tab: new spectra}. The spectra were obtained using a blue and a red arm with the 600 and 316 lines/mm grating respectively, 1.5" or 2" slit widths and 1x1 binning, and processed using standard procedures. All the observed sources fall within the redshift range 0.04$\leq z \leq$ 0.22, therefore covering the H~$\alpha$ for all sources. In some of the cases, the H~$\beta$ and [O~\textsc{iii}] emission lines were in the border between the  blue (2500 -- 5700 \AA) and red (4800 -- 10700 \AA) useful regions, leading to uncertain fits. Thus, we compared the broad H~$\alpha$ and the H~$\alpha$/[S \textsc{ii}] ratio between epochs, instead of the broad H~$\beta$ or the H~$\alpha$/[O \textsc{iii}] ratio as performed in other studies \citep[e.g. in ][]{graham2020}.



\subsection{Spectroscopic fitting}
We fit the archival DR17 SDSS spectra and the second epoch spectra from the Keck I and P200 telescopes using the Penalized Pixel-Fitting (pPXF) software \citep{cappellari2017improving}. To account for the stellar continuum component we used the E-MILES library \citep{vazdekis2010evolutionary}, and to model the AGN emission we added the following components:
\begin{itemize}
\item A power low template for the accretion disc contribution of the form $(\frac{\lambda}{\lambda_N})^\alpha$ where $\lambda$ is the wavelength, $\lambda_N= 5000$ \AA\ is a normalization factor and $\alpha$ goes from $-3$ to 0 in steps of 0.1.
\item One kinematic component with both permitted and forbidden emission lines, with free normalizations, to model the narrow lines. 
\item  One kinematic component with permitted emission lines, with free normalizations, to model the possible BELs.
\item One component for possible outflows with velocity dispersion values from 400 to 1000 km s$^{-1}$. 
\end{itemize}

To fit the second epoch spectra we used the same stellar population templates obtained from the SDSS fit and we left the normalization free during the fitting process. We obtained errors for each spectrum by performing Monte Carlo simulations using the best-fitting model and simulating random noise generated from the standard deviation of the best-fitting residuals. 
Then we fit the simulated spectra using the same procedure described for the SDSS and second epoch spectra, providing an error on the model parameters.

Table~\ref{tab: spectra} shows the results of the spectral fitting for broad H~$\alpha$ and the 1$\sigma$ error from the simulations. We identify 13 confirmed CL AGNs with >3$\sigma$ change in the EW of broad H~$\alpha$ \textit{and}  >3$\sigma$ change in the H~$\alpha$/[S \textsc{ii}] ratio. We also identify as CL two sources with a >3$\sigma$ change in the H~$\alpha$/[S \textsc{ii}] ratio, whose change can be confirmed via visual inspection. In total, we find 15 CL sources (highlighted in bold, see Table~\ref{tab: spectra}). Their optical spectra are shown in the Appendix, Fig.~\ref{fig: lc and spec}. Some of the sources show bluer continuum emission and/or asymmetric and complex BEL profiles (e.g., ZTF18acbzrll, ZTF19aavyjdn and ZTF20aaxwxgq) during the high state, as found in previous CL works \citep{2021Oknyansky}. 
In one case, ZTF19aafcyzr, the changes in the BELs according to the spectral fits are significant but they are not obvious when looking at the difference spectrum, so we speculate these changes could be driven by differences in the spectroscopy (that is, different instruments and set-up) and not to physical changes.
 
 For the 36 observed sources from our CL candidates, we distinguish the 19 confirmed CL AGNs as the CL sample (including the four CLs reported in \citetalias{2022lopeznavas}) and the 17 not confirmed CLs as the NOT CL sample (including two  such sources reported in \citetalias{2022lopeznavas}).
 
\begin{table*}
	\centering
	    \begin{adjustbox}{max width=\textwidth}
	    \begin{threeparttable}
	\caption{Sample of the CL candidates observed and results of the spectral fitting. $z$ denotes redshift from the DR17 SDSS data base. MJD, fiberid and plate denote the date and observational information from the SDSS spectra. EW denotes the equivalent width of the BELs. Sources identified as CL are shown in bold. Asterisks denote CL sources showing a >3$\sigma$ change in the H~$\alpha$/[S \textsc{ii}] ratio, but with no significant changes in EW H~$\alpha$. Some sources have two different ZTF IDs, which are separated by a slash. For completeness, we include the information for the six sources from \citetalias{2022lopeznavas} at the end of the table}
	\label{tab: spectra}

	\begin{tabular}{ccccccccccc} 
ZTF ID & RA & DEC & $z$ & MJD & fiberid & plate & EW H~$\alpha$ SDSS & EW H~$\alpha$ new & H~$\alpha$/[S \textsc{ii}] SDSS & H~$\alpha$/[S \textsc{ii}] new \\
& $\deg $& $\deg$ & & & & & \AA & \AA  \\
\hline
ZTF18aaiescp & 207.21292 & 57.646792 & 0.13 & 52668 & 198 & 1158 & 30$ \pm $2 & 14$ \pm $1& 1.75$ \pm $0.12& 1.14$ \pm $0.07\\
ZTF18aaiwdzt & 199.48361 & 49.258651 & 0.09 & 52759 & 390 & 1282 & 22$ \pm $2 & 24$ \pm $2& 0.7$ \pm $0.07& 0.69$ \pm $0.04\\
\textbf{ZTF18aajywbu} & 205.45327 & 37.013091 & 0.20 & 53858 & 513 & 2101 & 23$ \pm $4 & 94$ \pm $2& 2.4$ \pm $0.4& 9.01$ \pm $0.23\\
\textbf{ZTF18aaqftos} & 180.95505 & 60.888181 & 0.07 & 52405 & 420 & 954 & 23$ \pm $3 & 116$ \pm $3& 1.81$ \pm $0.21& 25.09$ \pm $0.74\\
ZTF18aaqjyon & 180.42264 & 38.47264 & 0.06 & 53473 & 358 & 2108 & 6$ \pm $2 & 0$ \pm $0& 0.56$ \pm $0.18& 0.0$ \pm $0.25\\
\textbf{ZTF21aaqlazo/ZTF18aasudup} & 170.03619 & 34.312731 & 0.04 & 53713 & 346 & 2100 & 15$ \pm $1 & 43$ \pm $1& 1.79$ \pm $0.11& 6.43$ \pm $0.2\\
\textbf{ZTF18aavxbec}* & 243.08151 & 46.495172 & 0.13 & 52370 & 408 & 814 & 17$ \pm $2 & 16$ \pm $0& 1.47$ \pm $0.18& 3.14$ \pm $0.1\\
\textbf{ZTF18aawoghx} & 156.10498 & 37.650863 & 0.10 & 52998 & 254 & 1428 & 9$ \pm $2 & 34$ \pm $1& 0.91$ \pm $0.2& 3.95$ \pm $0.17\\
\textbf{ZTF18aawwcaa} & 128.10120 & 35.859979 & 0.14 & 52668 & 459 & 1197 & 26$ \pm $2 & 73$ \pm $2& 5.49$ \pm $0.77& 16.74$ \pm $1.06\\
ZTF19aabyvtv/ZTF18aayyapb & 197.87773 & 31.866893 & 0.07 & 53819 & 632 & 2029 & 17$ \pm $1 & 6$ \pm $3& 17.57$ \pm $3.38& 8.09$ \pm $9.03\\
\textbf{ZTF18acbzrll} & 124.82294 & 30.32660 & 0.10 & 52619 & 94 & 931 & 7$ \pm $2 & 90$ \pm $3& 1.29$ \pm $0.43& 20.13$ \pm $0.88\\
ZTF18acgvmzb/ZTF18aclfugf & 148.96896 & 35.965616 & 0.04 & 52998 & 251 & 1596 & 8$ \pm $1 & 8$ \pm $1& 0.52$ \pm $0.03& 0.73$ \pm $0.04\\
ZTF18achdyst & 157.39925 & 24.777606 & 0.11 & 53734 & 545 & 2349 & 26$ \pm $1 & 15$ \pm $1& 5.72$ \pm $0.02& 1.37$ \pm $0.1\\
ZTF18acusqpt/ZTF18adppkkj & 177.91898 & 12.036714 & 0.07 & 53142 & 116 & 1609 & 18$ \pm $1 & 17$ \pm $2& 3.27$ \pm $0.32& 6.07$ \pm $0.65\\
\textbf{ZTF19aafcyzr} & 125.71008 & 15.673859 & 0.12 & 53713 & 517 & 2272 & 11$ \pm $1 & 30$ \pm $1& 1.2$ \pm $0.17& 3.98$ \pm $0.24\\
ZTF19aaixgoj & 146.80538 & 12.205624 & 0.12 & 53053 & 575 & 1742 & 14$ \pm $1 & 13$ \pm $1& 2.65$ \pm $0.29& 2.61$ \pm $0.24\\
\textbf{ZTF19aaoyjoh} & 180.18956 & 14.967685 & 0.11 & 53463 & 158 & 1763 & 10$ \pm $5 & 59$ \pm $1& 1.25$ \pm $0.37& 10.44$ \pm $0.31\\
ZTF19aapehvs & 199.74519 & 57.501847 & 0.10 & 52759 & 338 & 1320 & 32$ \pm $2 & 18$ \pm $1& 3.02$ \pm $0.27& 2.34$ \pm $0.17\\
ZTF19aavqrjg & 181.24583 & 15.58718 & 0.22 & 53467 & 447 & 1764 & 51$ \pm $3 & 67$ \pm $9& 35.65$ \pm $2.0& 10.61$ \pm $10.55\\
\textbf{ZTF19aavyjdn} & 202.39941 & $-$1.509453 & 0.08 & 52377 & 524 & 910 & 14$ \pm $4 & 77$ \pm $18& 2.94$ \pm $0.63& 31.1$ \pm $6.41\\
ZTF20aaeutuz & 164.81581 & 12.483378 & 0.15 & 55956 & 726 & 5357 & 30$ \pm $1 & 34$ \pm $1& 2.83$ \pm $0.11& 5.31$ \pm $0.2\\
ZTF20aagwxlk & 153.42829 & 55.432205 & 0.15 & 52407 & 179 & 946 & 31$ \pm $2 & 27$ \pm $8& 3.69$ \pm $0.29& 1.14$ \pm $0.64\\
\textbf{ZTF20aagyaug} & 172.41282 & 36.883602 & 0.20 & 55673 & 426 & 4648 & 41$ \pm $3 & 96$ \pm $2& 1.35$ \pm $0.09& 4.17$ \pm $0.11\\
ZTF20aakreaa & 181.52333 & 42.169888 & 0.10 & 53120 & 149 & 1448 & 17$ \pm $1 & 20$ \pm $10& 4.85$ \pm $0.92& 1.03$ \pm $2.34\\
ZTF20aaorxzv & 132.39052 & 3.68048 & 0.08 & 52224 & 606 & 564 & 18$ \pm $1 & 16$ \pm $16& 1.63$ \pm $0.16& 1.04$ \pm $0.31\\
\textbf{ZTF21abwoxbv/ZTF20aaxwxgq} & 234.63610 & 46.126392 & 0.20 & 52781 & 559 & 1332 & 11$ \pm $3 & 68$ \pm $10& 2.47$ \pm $0.74& 56.73$ \pm $9.03\\
\textbf{ZTF20abcvgpb}* & 238.24977 & 21.046358 & 0.17 & 53557 & 220 & 2171 & 49$ \pm $4 & 44$ \pm $1& 2.98$ \pm $2.3& 11.77$ \pm $0.69\\
ZTF20abgnlgv & 232.52715 & 7.172269 & 0.13 & 54208 & 322 & 1820 & 41$ \pm $2 & 35$ \pm $1& 3.98$ \pm $0.38& 4.8$ \pm $0.17\\
\textbf{ZTF21aafkiyq} & 174.93478 & $-$1.727439 & 0.11 & 52294 & 614 & 327 & 18$ \pm $2 & 54$ \pm $1& 1.79$ \pm $0.23& 4.42$ \pm $0.17\\
\textbf{ZTF21abcsvbr} & 184.64829 & 18.771713 & 0.22 & 54477 & 299 & 2611 & 19$ \pm $4 & 44$ \pm $1& 1.37$ \pm $0.25& 5.35$ \pm $0.29\\
\hline
\hline

\textbf{ZTF19abixawb} & 2.56204 & 0.13912 & 0.10 & 51793 &516& 388& 8$^{+3}_{-1}$ &47.0$^{+10}_{-0.3}$ & &\\
\textbf{ZTF20abshfkf}& 18.29924 & 1.59516 &  0.24 & 57282 &83& 7859 &7$^{+20}_{-4}$&133$\pm$2&\\

		\textbf{ZTF18accdhxv}& 118.93488 & 19.39342  &0.11 & 53315& 306 &1922&21$^{+9}_{-1}$ &90$^{+3}_{-20}$ \\
  
		\textbf{ZTF19aalxuyo} & 123.16978 & 7.25791  & 0.08 & 53083 & 294&1757&  2.2$^{+5}_{-0.4}$&54$^{+1}_{-3}$& & \\

		ZTF19aaxdiui & 325.19182 & 9.27550 &  0.40 & 55475 &234 &4093 & -- & -- && \\

		ZTF18abtizze & 327.73225 & -1.11506 &  0.09 & 53172	& 138 & 1031 & 3$^{+21}_{-1}$ &13$^{+1}_{-4}$\\

    \hline
	\end{tabular}
	\end{threeparttable}
	\end{adjustbox}
\end{table*}

\section{Results}

\subsection{Improvement of the selection method: ALeRCE features for the alert light curves}\label{section: improvement}
All the CL candidates considered in this work have generated at least 6 ZTF alerts in either \textit{g} or \textit{r} band and have been classified by the AleRCE LCC. An alert is generated when a 5$\sigma$ variation in the template-subtracted image occurs. In this section, we analyse the variability of the sample to determine whether the CL phenomenon is related to any physical parameter or if we can make other improvements to the selection method and thereby to the CL candidate list.

The LCC uses a total of 174 features, most of them computed solely with the public ZTF \textit{g} and \textit{r} data. The complete set of features is described in the ALeRCE website\footnote{\url{http://alerce.science/features/}}, and can be requested using the ALeRCE \textsc{python} client. In this work, we separate the features that dominate the classifier (both the `Top' level and the `Stochastic' level of the LCC) as reported in \citet{sanchez2021} and the secondary, not-ranked features. For comparison, we obtained the features for the known AGNs that were used to train the LCC, which includes the Weak Type 1 sources from \citet{Oh2015} and the host-dominated AGNs (class `A') from MILLIQUAS, totalling 4612 sources.


\subsubsection{Top-ranked variability and colour features}
Most of the features that dominate the LCC consist of ZTF and AllWISE colours and variability features related to the amplitude and time-scale of the variability and to a decrease/increase of the luminosity. In order to evaluate the difference in distribution of these features between the CL and NOT CL samples we applied the Kolmogorov--Smirnov (KS) test to all their ranked-features. In Table \ref{tab: features} we present the features that dominate the LCC and have a p-value <0.05, that is, where we can reject the null hypothesis that the two distributions (from the CL and the NOT CL samples) are identical.

\begin{table*}
	\centering
	    \begin{adjustbox}{max width=\textwidth}
	    \begin{threeparttable}
	\caption{Features that show different distributions for the CL and the NOT CL samples. Features recovered for both the alert light curves and the forced-photometry light curves are highlighted in bold.   }
	\label{tab: features}

	\begin{tabular}{lcl} 
\textbf{Name}&\textbf{Filter}&\textbf{Description}\\
\hline
\multicolumn{3}{c}{Top ranked variability and colour features used in the LCC}\\
\hline

\textbf{SF$\_$ML$\_$amplitude} & \textit{g} & rms magnitude difference of the structure function computed over a one year time-scale\\ 
\textbf{MHPS$\_$low}  & \textit{g} & Variance associated with the 100 days time-scale ('low' frequency)\\
\textbf{SPM$\_$tau$\_$rise} & \textit{g} &  Initial rise time-scale from the supernova parametric model  	\\ 
\textbf{GP$\_$DRW$\_$sigma}  &g &Amplitude of the variability at short time-scales, from the damp random walk (DRW) model 	\\ 
GP$\_$DRW$\_$tau&g	&Relaxation time  from the DRW model\\ 
IAR$\_$phi&g& Level of autocorrelation from an irregular autoregressive (IAR) model	\\ 
positive$\_$fraction & \textit{g},\textit{r} & Number of detections in the difference images that are brighter than the template \\
delta$\_$mag$\_$fid& \textit{g}	& Difference between the maximum and minimum observed magnitudes in a given band\\ 
\textit{r}--\textit{\textit{W3}} & & colour computed using the ZTF mean \textit{r} magnitude and the AllWISE \textit{\textit{W3}} filter\\
g$\_$r$\_$mean$\_$corr & & ZTF \textit{g}--\textit{r} colour using the mean magnitudes of each band \\
g$\_$r$\_$max & & ZTF \textit{g}--\textit{r} colour using the brightest magnitudes of each band\\

\hline
\multicolumn{3}{c}{Other features used in the LCC}   \\
\hline
n$\_$pos & \textit{g},\textit{r} & Number of positive detections in the alert light curve \\
n$\_$neg & \textit{g},\textit{r} & Number of negative detections in the alert light curve \\
iqr &\textit{g}& Difference between the 3rd and the 1st quartile of the light curve \\
\textbf{MHPS$\_$ratio}&\textit{g}&  Ratio between the variances at 100 days and 10 days time-scales for a given band, applying a Mexican hat filter 	\\ 

	\hline
	\multicolumn{3}{c}{Variability features for the forced-photometry light curves} \\
	\hline

SPM$\_$A   & \textit{r} &  Amplitude from the supernova parametric model \\
LinearTrend  & \textit{g},\textit{r} & Slope of a linear fit to the light curve \\
ExcessVar  &\textit{g}& Intrinsic variability amplitude \\

Meanvariance  &\textit{g}&Ratio of the standard deviation to the mean magnitude\\
Std  &\textit{g}&Standard deviation of the light curve \\
Amplitude &\textit{g}& Half of the difference between the median of the maximum 5 per cent and of the minimum 5 per cent magnitudes \\
\textbf{SPM$\_$tau$\_$rise}   &\textit{g}, \textit{r} & See above \\
\textbf{MHPS$\_$low} & \textit{r} & See above\\
\textbf{MHPS$\_$ratio} & \textit{r} & See above\\
\textbf{GP$\_$DRW$\_$sigma}  & \textit{g},\textit{r} & See above\\ 
\textbf{SF$\_$ML$\_$amplitude}  & \textit{g},\textit{r} & See above\\

    \hline
	\end{tabular}
	\end{threeparttable}
	\end{adjustbox}
\end{table*}

We note that the DRW parameters determination is generally biased for light curve lengths shorter than 10 times the true $\tau$ value \citep{2017koz,sanchez2017}, which is the case of our ZTF data. Therefore, the DRW parameters obtained in this work should be considered just as variability features and not as physically correct estimations.  

\begin{figure*}
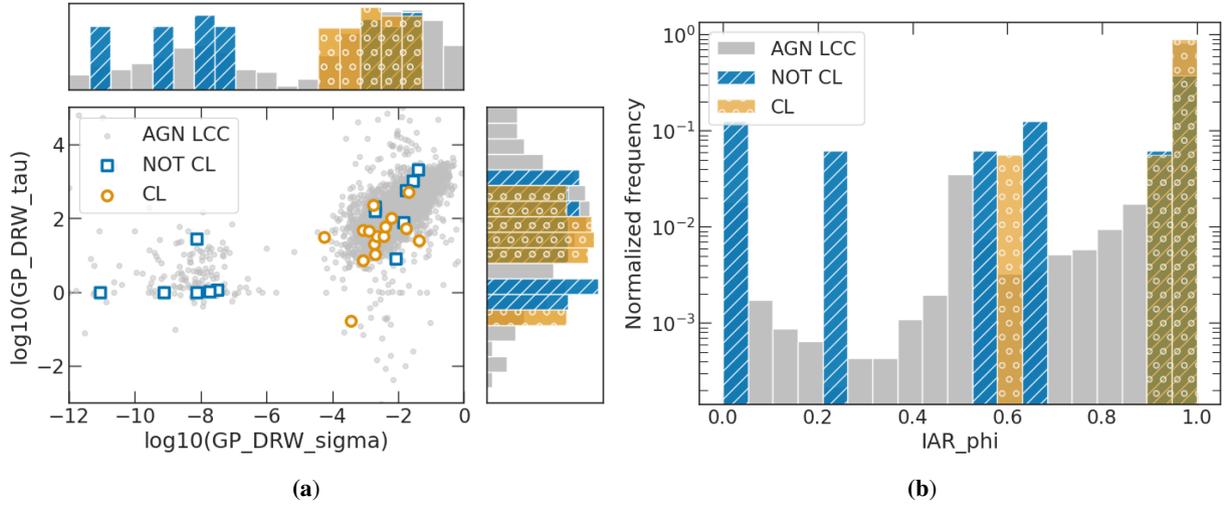

        \centering
        \begin{subfigure}{0.45\textwidth} 
            \includegraphics[width=\textwidth]{plots/sigma_tau.pdf}
            \caption{\normalsize{\textbf{(a})}}
            \label{fig:sigmatau}
        \end{subfigure}
        \begin{subfigure}{0.45\textwidth} 
            \includegraphics[width=\textwidth]{plots/iarphi.pdf}
            \caption{\normalsize{\textbf{(b})}}
            \label{fig:var_iarphi}
        \end{subfigure}

     \caption{Distributions of the alert light curves top-ranked variability features for the CL and NOT CL samples and the AGN LCC training set: (a) relaxation time and amplitude of the variability on short time-scales, obtained from the damp random walk (DRW) model;  and (b) level of autocorrelation from an irregular autoregressive (IAR) model. These features show distinct distributions between the CL and NOT CL samples and could be used to select the most promising CL candidates.}
\label{fig:var_feat}
\end{figure*}

In Fig.~\ref{fig:var_feat} we show the distribution of some of the variability features computed in the \textit{g} band that present different distributions between the CL and NOT CL sources. In particular, the DRW relaxation time for the NOT CL objects peaks at the minimum data sampling ($\sim$ 1 d), and spreads up to >1000 d, larger than the maximum light curve length. This indicates a DRW model is unable to properly model the optical variability for some NOT CLs, and thus it is unlike Type-1 AGN. For the CLs however, the DRW relaxation time peaks at 10--100 days as expected for Type 1 AGNs. In terms of the amplitude of the variability, from the GP$\_$DRW$\_$sigma distributions we see that some NOT CL objects reach much smaller values (log10(GP$\_$DRW$\_$sigma) $ < -6$), which again indicates they have most likely been misclassified as Type 1 AGN by the LCC. Interestingly, the amplitude of the variations for all our objects peaks at a smaller value than the distribution for the AGN training set, suggesting their variability could be diluted by the host galaxy contribution. On the other hand, the autocorrelation of the light curves, given by the IAR$\_$phi parameter, reaches smaller values for the NOT CL sample than for the CL sample. These features could be used to further clean the CL candidate list.

\begin{figure*}
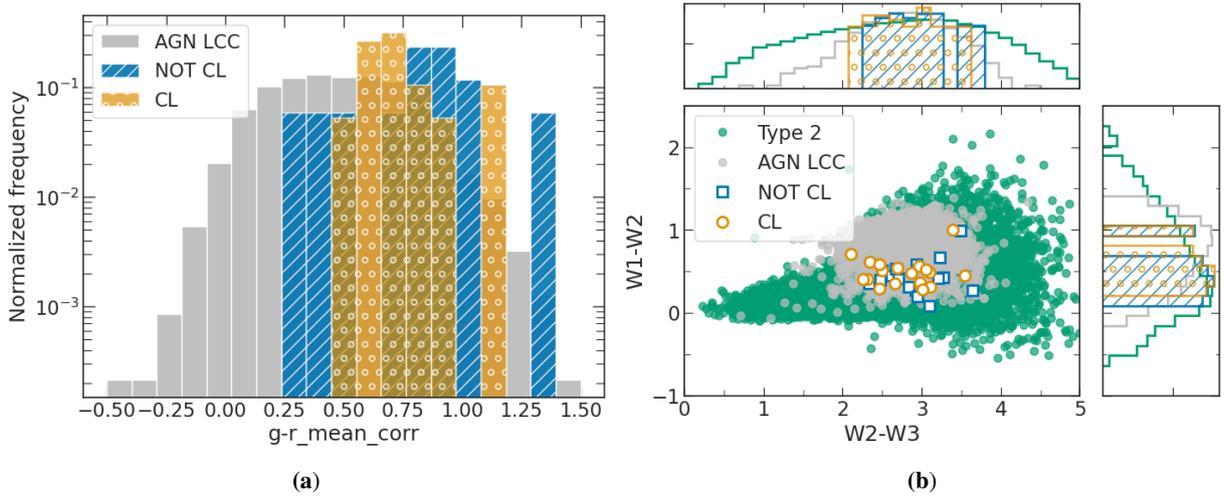

        \centering
            \begin{subfigure}{0.45\textwidth} 
            \includegraphics[width=\textwidth]{plots/g-r-mean-corr.pdf}
            \caption{\normalsize{\textbf{(a})}}
            \label{fig:g-r}
        \end{subfigure}    
   \begin{subfigure}{0.45\textwidth} 
            \includegraphics[width=\textwidth]{plots/allwise.pdf}
            \caption{\normalsize{\textbf{(b})}}
            \label{fig:allwise}
        \end{subfigure}  
     \caption{Alert light curves top-ranked colour features for the CL and NOT CL samples and the AGN LCC training set: (a) \textit{g--r} colour obtained with ZTF data and (b) AllWISE MIR colours in comparison with the Type 2 parent sample. Both the CL and NOT CL samples show optical and MIR colours more host-galaxy dominated than the AGN LCC training set.} 
\label{fig:other_feat}
\end{figure*}

Apart from the variability features, the classifier is also dominated by ZTF and AllWISE colours and the morphological  properties  of  the  images. In general, the optical colours for both samples look similar to each other but show a redder tendency than the AGN training set distribution, as shown in Fig.~\ref{fig:g-r}. 

In Fig.~\ref{fig:allwise} we present the 2010--2011 AllWISE  \textit{W1}--\textit{W2}  versus the  \textit{W2}--\textit{W3} colours for the CL and NOT CL objects, in comparison to the Type 2 parent sample and the AGN training set. Most of our objects have fairly similar AllWISE colours, but the distribution of the  \textit{W1}--\textit{W2}  colour peaks at a lower value than the AGN training set and closer to the Type 2 distribution, which implies that the MIR colours are generally dominated by the stellar populations. 
We note that the AllWISE observations were taken between 2010 and 2011, so these features are not indicative of their \textit{current} MIR colour. We further investigate the behaviour of the MIR colours in Section \ref{sect:wise}, including contemporaneous \textit{WISE} observations.  

\begin{figure}

\includegraphics[width=0.45\textwidth]{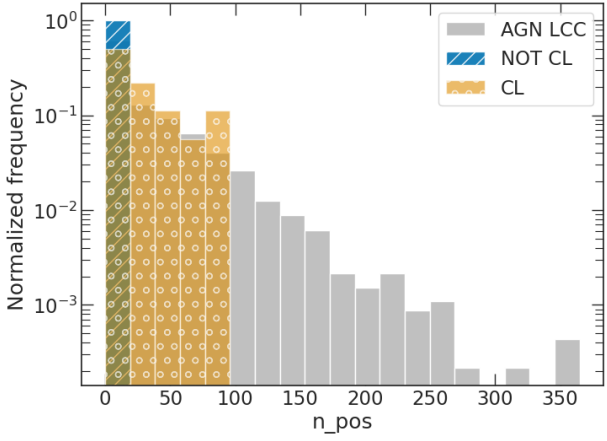}
    \caption{Number of positive detections in the alert light curves for the CL and NOT CL samples and the AGN LCC training set, where the CL sample tends to increase their optical flux.}
    \label{fig:npos}

\end{figure}    
        
\subsubsection{Not-ranked features}
The top-ranked features are the most important features that the LCC uses for the classification of variable sources. However, there are secondary features that could potentially allow us to evaluate whether a source is a good CL candidate or not. In Table \ref{tab: features} we present other secondary features showing distinct distributions between the CL and NOT CL samples. A particular example is the number of positive detections in the alert light curves (n$\_$pos, see Fig.~\ref{fig:npos}), which reaches higher values for the CL sample, meaning that CL objects tend to increase more their flux with respect to the template image (as expected for turning-on events). 



\subsection{Characteristics of the CL vs NOT CL AGNs}

\subsubsection{BPT diagnostics}
\begin{figure}
    \includegraphics[width=0.45\textwidth]{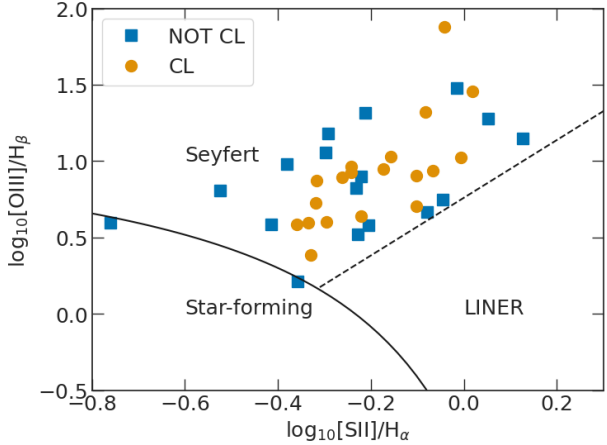} 
    \caption{BPT diagram for the CL and NOT CL samples obtained from their archival SDSS spectra. The solid and dashed lines show the classification scheme from \citet{2001Kewley, 2006kewley}. The comparison indicates these samples are indistinguishable in terms of the BPT diagnostic criteria. }
\label{fig: bpt}
\end{figure}

To investigate the emission-line properties of the samples, we calculated the BPT \citep{1981BPT} diagnostics from the archival SDSS spectra. We used the classification system defined by \citet{2006kewley} for different ionisation mechanisms utilizing the three BPT diagnostic criteria ([N \textsc{ii}], [S \textsc{ii}], and [O \textsc{i}]) and found all sources are consistent with a Seyfert classification. The KS-test leads to large p-values >0.4 for the three cases, showing that these samples are indistinguishable in terms of the emission lines properties from their old (pre-CL) optical spectra. Similar results were found by analysing second epoch spectra. As an example, the [S \textsc{ii}] BPT diagram for the CL and NOT CL samples is plotted in Fig.~\ref{fig: bpt}. 

\subsubsection{Eddington ratio estimates}
\begin{figure*}
    \includegraphics[width=0.7\textwidth]{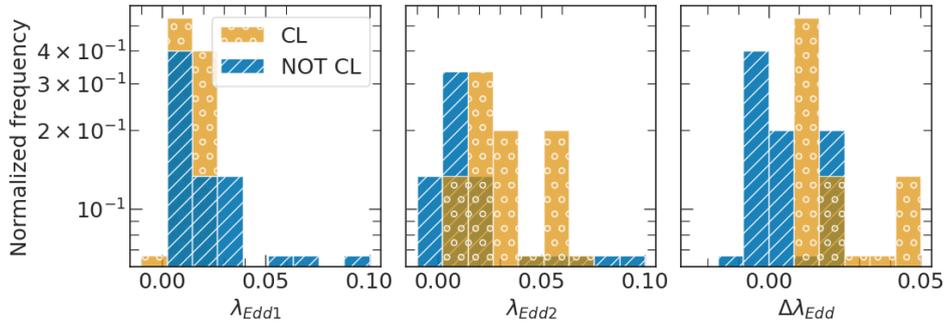} 
    \caption{Eddington ratios for the CL and NOT CL samples obtained from the archival SDSS spectrum (left panel) and current spectra (middle panel), and their difference ( $\Delta \lambda_{\rm   \rm Edd}$ = $\lambda_{\rm   \rm Edd2} - \lambda_{\rm   \rm Edd1}$, right panel). CL objects have increased their Eddington ratios and are now accreting at 1--5 per cent $L_{\rm Edd}$. }
\label{fig: edd}
\end{figure*}

We estimated the black hole masses ($M_{\rm BH}$) and continuum luminosity at 5100 \AA~($L_{5100}$) using the full width at half maximum (FWHM) and luminosity of broad H~$\alpha$ as outlined in \citet{reines2013}, using the values obtained from the new spectra. Then, we computed the Eddington ratios for the old and new spectra $\lambda_{\rm   \rm Edd}=L_{\rm bol}/L_{\rm Edd}$, where $L_{\rm Edd} = 1.5 \times 10^{38}(M_{\rm BH} / $M$_\odot)$ erg s$^{-1}$ is the Eddington luminosity and  $L_{\rm bol}$ is the bolometric luminosity defined as $L_{\rm bol}=40 \cdot (L_{5100}/10^{42})^{-0.2}$ erg s$^{-1}$ according to \citet{2019netzer}. In Fig.~\ref{fig: edd} we present $\lambda_{\rm   \rm Edd}$ for both samples computed for the old ($\lambda_{\rm   \rm Edd1}$) and new spectra ($\lambda_{\rm   \rm Edd2}$), and the difference of accretion rate ($\Delta \lambda_{\rm   \rm Edd}$). We find that the old accretion rate is similar for both samples, but in the second epoch spectra the distribution shifts towards higher values for the CLs. These are the expected results for turning-on AGNs, since both the method to compute $\lambda_{\rm   \rm Edd}$ and the criteria to confirm the sources as CL (e.g. >3$\sigma$ change in the EW of broad H~$\alpha$) use the properties of broad H~$\alpha$.

\subsubsection{ALeRCE features for the forced-photometry light curves}
\begin{figure}
        \centering
            \begin{subfigure}{0.45\textwidth} 
            \includegraphics[width=\textwidth]{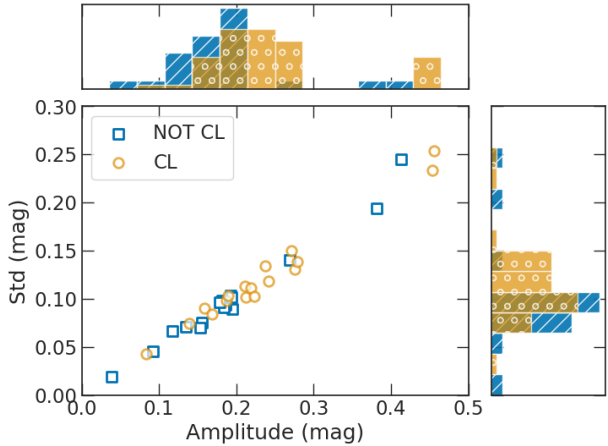}
        \end{subfigure}    

     \caption{Amplitude and  standard deviation in the \textit{g} filter for the CL and NOT CL forced-photometry light curves. CL objects present higher values for both features, indicating a stronger optical variability.} 
\label{fig:forced}
\end{figure}

To compare with the alert light curves, we also analysed variability in the full forced-photometry light curves, which are produced based on all ZTF difference images available. We requested the most updated forced photometry (up to 2022 September 26) of the entire sample from the ZTF forced-photometry service and generated the cleaned light curves according to the recommendations outlined in \citet{masci2019}.  A \textsc{python} library to extract variability features in astronomical light curves is publicly available\footnote{\url{https://github.com/alercebroker/lc_classifier}}.  The forced-photometry light curves have a mean of 376 (453) data points in the \textit{g} (\textit{r}) filter, in comparison with the 30 (27) detections from the alert light curves. As a result, the variability features can be better constrained and we find more top-ranked features that show different distributions according to the KS-test between the CL and NOT CL samples, as shown in Table \ref{tab: features}. We recover many variability features related to the amplitude of the variations and the deviations from the mean (e.g. ExcessVar, Meanvariance or Std), which are missing in the comparison of features from the alert light curves. Fig.~\ref{fig:forced} shows the amplitude and the standard deviation in the \textit{g} filter for the forced-photometry light curves for both samples. The CL objects present generally higher values for both features, indicating that their variability is more similar to the expected Type 1 behaviour than the NOT CL sample. 

\subsubsection{Mid-infrared variability}\label{sect:wise}
\begin{figure}
        \centering
            \begin{subfigure}{0.4\textwidth} 
            \includegraphics[width=\textwidth]{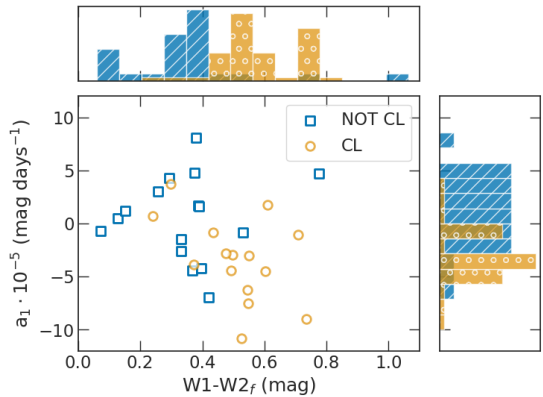}
        \end{subfigure}

     \caption{Distributions of the linear slope of the  \textit{W1}  10-years-long light curve (a1) and the last epoch colour (\textit{W1}--\textit{W2}$_f$) for the CL and NOT CL samples. The lower a1 values for the CL sample indicates the sources are getting brighter in the MIR waveband, while the slopes for the NOT CL distribute around zero, indicating that as a sample they are neither brightening nor dimming. The higher last epoch colour (\textit{W1}--\textit{W2}$_f$) for the CLs is expected for AGN-dominated galaxies (\textit{W1}--\textit{W2} >0.5). }
\label{fig: mir}

\end{figure}

Some of the features that the LCC uses to classify variable objects are computed with AllWISE data, which are indicative of the state of the sources ten years ago. To investigate whether the CL and NOT CL samples show distinct MIR behaviour we downloaded all the AllWISE multi-epoch and NEOWISE-R single exposure (L1b) photometric data spanning from 2010 to 2021 and averaged every six months. For each source, we obtained the following variability features for the \textit{W1} and  \textit{W2} bands: the maximum magnitude and colour variations ($\Delta$\textit{W1}, $\Delta$\textit{W2}, $\Delta$\textit{W1}--\textit{W2}), the colour from the last epoch (\textit{W1}--\textit{W2}$_f$), the intrinsic variability ($\sigma_{m1}$ and $\sigma_{m2}$) computed as in \citet{2022lyu} and the slopes of a linear trend fit to the MIR magnitude light curves (a$_1$ and a$_2$) and to the  \textit{W1}--\textit{W2}  colour (a$_{12}$). Table \ref{tab:mir} shows the comparison between the median values with the 1$\sigma$ errors for the CL and NOT CL samples. All the features except for the maximum colour variation $\Delta$\textit{W1}--\textit{W2}  show distinct distributions according to the KS-test, with CL sources having a stronger variability. Moreover, the results from the linear fits indicate that the CLs have become brighter in both bands and have higher  \textit{W1}--\textit{W2}  values (see Fig.~\ref{fig: mir} and \ref{fig: lc and spec}), whereas for the NOT CL the distributions peak closer to zero resulting in no net increase or decrease in brightness or colour.    

\begin{table}
	\centering
	    \begin{adjustbox}{max width=0.7\textwidth}
	    \begin{threeparttable}
	\caption{Mid-infrared variability features. Asterisks indicate the features that show distinct distributions between the samples according to the KS-test (p-value<0.05). The errors correspond to the 1$\sigma$ deviation from the median. }
	\label{tab:mir}

	\begin{tabular}{ccc} %
\textbf{Feature}&\textbf{CL}&\textbf{NOT CL}\\
\hline
$\langle\Delta$\textit{W1}$\rangle$ (mag)*&0.5$^{+0.3}_{-0.1}$&0.4 $\pm$0.2 \\
$\langle\Delta$\textit{W2}$\rangle$ (mag)*&0.7$\pm$0.3&0.5$^{+0.1}_{-0.3}$ \\
$\langle\Delta$\textit{W1}--\textit{W2} $\rangle$(mag)&0.3$^{+0.2}_{-0.1}$&0.3$^{+0.2}_{-0.1}$\\
$\langle$\textit{W1}--\textit{W2}$_{f}\rangle$(mag)*&0.5$^{+0.2}_{-0.1}$&0.4$^{+0.1}_{-0.2}$\\
$\langle\sigma_{m1}\rangle$ *&0.15$\pm$0.05& 0.09$\pm$0.07 \\
$\langle\sigma_{m2}\rangle$*&0.18$^{+0.09}_{-0.05}$&0.14$^{+0.04}_{-0.09}$ \\
$\langle$a$_1\rangle \cdot 10^{-5}$ (mag$^{-1}$)*&$-4^{+4}_{-9}$&0.5$^{+4}_{-5}$ \\
$\langle$a$_2\rangle\cdot 10^{-5}$ (mag$^{-1}$)*&$^{+9}_{-13}$&2$^{+4}_{-8}$\\
$\langle$a$_{12}\rangle\cdot 10^{-5}$(mag$^{-1}$)*&2$^{+3}_{-4}$&0$^{+3}_{-2}$ \\

\hline
    \end{tabular}
    \end{threeparttable}
    \end{adjustbox}
\end{table}

\subsubsection{X-ray variability}
In order to obtain the X-ray fluxes of our sources, we have used the individual eROSITA \citep[extended ROentgen Survey with an Imaging Telescope Array, ][]{2021erosita} All-Sky Surveys (eRASS1 to eRASS5). The data were processed with the eROSITA Standard Analysis Software System (eSASS, \citealt{2022brunner}). We used the newest available pipeline processing version c020 which is an updated version of the software used for the first eROSITA Data Release (Merloni et al. 2023, in prep).

The counterparts are determined using the same procedure adopted in the eROSITA/eFEDS field \citep{2022salvato}, but applied to Legacy Survey DR10\footnote{\url{https://www.legacysurvey.org/dr10/}} and Gaia DR3 separately. After the identification of the CL candidates with the counterparts, we obtained the 0.2--2.3 keV flux from the corresponding eROSITA catalog (see \citealt{2022brunner} for a description of the eROSITA catalog processing).


%

\begin{figure}
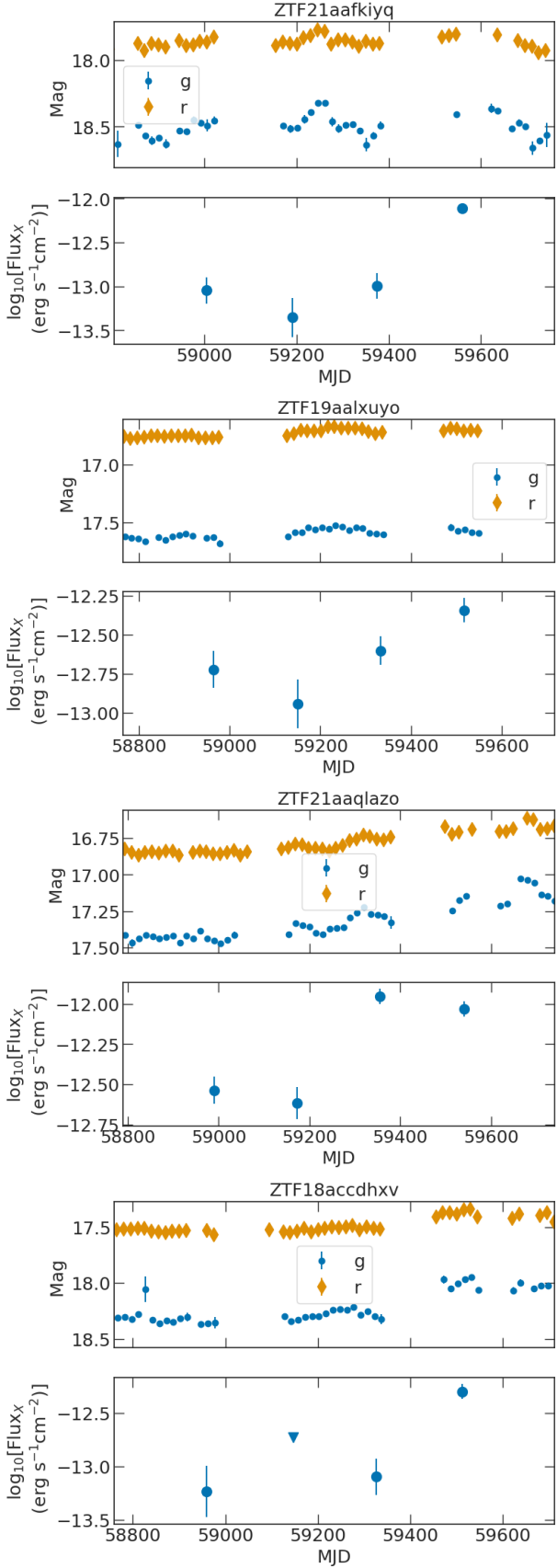

        \centering
            \begin{subfigure}{0.45\textwidth} 
                \includegraphics[width=\textwidth]{plots/incr1.pdf}
            \end{subfigure}    
            \begin{subfigure}{0.45\textwidth} 
                \includegraphics[width=\textwidth]{plots/incr2.pdf}
            \end{subfigure}   
            \begin{subfigure}{0.45\textwidth} 
                \includegraphics[width=\textwidth]{plots/incr4.pdf}
            \end{subfigure}   
            \begin{subfigure}{0.45\textwidth} 
                \includegraphics[width=\textwidth]{plots/incr3.pdf}
            \end{subfigure}
        \caption{ZTF forced-photometry light curves and contemporaneous eROSITA fluxes in the 0.2--2.3 keV band. The triangle in the last plot indicates an upper limit. These sources experience an increase in their X-ray flux during the eROSITA monitoring.} 
\label{fig: xraylc}
\end{figure}

From our list of 61 CL candidates, there are 28 sources within the eROSITA-DE footprint (Galactic longitude 179.9442 < $l$ < 359.9442 deg): 11 CLs, seven NOT CLs and ten CL candidates without a second epoch optical spectrum. From the CLs, ten sources have at least one detection within the five different eRASS, and one (ZTF18aawoghx) has only upper limits. The upper limits are calculated based on X-ray photometry on the eROSITA standard pipeline data products (science image, background image, and exposure time) following the Bayesian approach described by \citet{1991kraft}. For details about eROSITA upper limits, see Tub\'in-Arenas et al. (2023, in prep). We consider a circular aperture with a radius given by a PSF encircled energy fraction of EEF = 0.75 ($\sim 30\arcsec$) and a single-sided 3$\sigma$ confidence level. From the NOT CLs, four sources have been detected in at least one eRASS. The remaining three (ZTF18acgvmzb/ZTF18aclfugf, ZTF19aaixgoj, and ZTF20aaorxzv) have only upper limits. Interestingly, six CL sources show an X-ray flux increase between eROSITA scans by factors of 2 to 15 times. For two sources (ZTF19aavyjdn and ZTF21abcsvbr) the difference between the maximum and the minimum values is similar to the error of the minimum value. For the remaining four sources (ZTF18accdhxv, ZTF19aalxuyo, ZTF21aafkiyq and ZTF21aaqlazo) the difference is at least eight times the error (see  Fig.~\ref{fig: xraylc}).



 We also checked archival X-ray fluxes from other missions to compare to the eROSITA fluxes. All the 11 CLs and 7 NOT CLs in the eROSITA-DE footprint have at least one X-ray upper limit from either the \textit{XMM-Newton} Slew \citep{2008xmm} or \textit{ROSAT} Survey \citep{2016rosat}. However, due to the low sensitivity of the data, most of the archival upper limits fall above the current eROSITA measurements. This hinders us from finding the possible changes, with the notable exception of four of the CLs that show a significant ($\gtrapprox 1.5 \sigma$) increase in the eROSITA 2021 flux with respect to the archival 1990--1993 \textit{ROSAT} 1$\sigma$ upper limits or fluxes, which are shown in Table \ref{tab:xraysarchival}. In the table, we converted the observed \textit{ROSAT} fluxes to the 0.2--2.3 keV band for a direct comparison, using an absorbed power law model with photon index $\Gamma=2$ and column density $N_{\rm H}=3\cdot 10^{20}$ cm$^{-2}$. These sources are also the CLs that experience a significant X-ray increase during the eROSITA monitoring as shown in Fig.~\ref{fig: xraylc}.

\begin{table}
	\centering
	
	\begin{tabular}{ccc} %
ZTF ID & \textit{ROSAT} flux  & eROSITA flux  \\
& $\cdot$ 10$^{-13}$ erg s$^{-1}$ cm$^{-2}$  & $\cdot$ 10$^{-13}$ erg s$^{-1}$ cm$^{-2}$  \\
\hline
ZTF18accdhxv& <2.853  & 5.0 $\pm$ 0.8\\ 
ZTF19aalxuyo& <3.279 & 4.6 $\pm$ 0.8 \\ 
ZTF21aafkiyq& <1.049  & 7.7 $\pm$ 0.9\\
ZTF21aaqlazo& 0.51 $\pm$ 0.06 & 11.1 $\pm$ 0.1\\
    \hline
    \end{tabular}
\caption{0.2--2.3 keV X-ray fluxes for four CL sources that show an increase between the archival 1990--1993 \textit{ROSAT} 1$\sigma$ upper limits or fluxes and the 2021 eROSITA data. }
\label{tab:xraysarchival}
\end{table}


\begin{figure*}
        \centering
            \begin{subfigure}{0.6\textwidth} 
            \includegraphics[width=\textwidth]{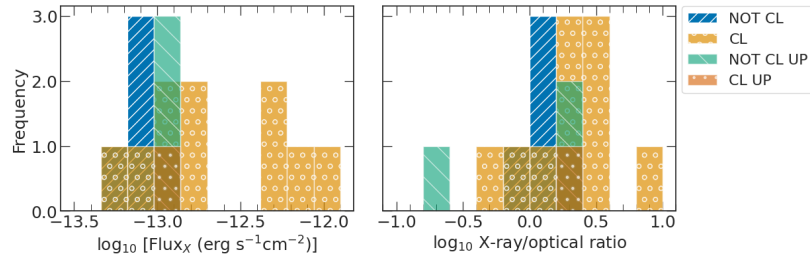}
        \end{subfigure}    
        
     \caption{Distributions of the 0.2--2.3 keV eROSITA fluxes (left) and ratios between the 0.2--2.3 keV flux and the \textit{g}-band flux from the ZTF forced photometry light curves, taken within the same day or week (right). \textit{UP} indicates the eROSITA upper limits.}
\label{fig: xray}
\end{figure*}

To compare the eROSITA fluxes between the CL and NOT CL sources, we selected the last X-ray detection within the five eRASS,
or the last eROSITA upper limit for the sources without detections. We also computed the ratio between the X-ray flux and the optical flux in the \textit{g} band, obtained from the ZTF forced photometry light curves. To avoid spurious results coming from variability, we chose pairs of contemporaneous fluxes, i.e., that were taken within the same days or week in the X-ray and optical bands. The results are plotted in Fig.~\ref{fig: xray}, which shows the CL sources are generally brighter in the X-ray band than the NOT CL sources, both in absolute terms and relative to their \textit{g} band fluxes. The KS-test indicates the X-ray flux distribution is significantly distinct between the CL and NOT CL samples (p-value $<0.05$), both considering just detections (p-value $=0.01$) and considering detections and upper limits (p-value $=0.01$). However, the X-ray to optical ratio distributions are not significantly distinct according to the KS-test, either considering just detections (p-value $=0.08$) or considering detections and upper limits (p-value $=0.10$). Therefore, although the X-ray to optical ratios tend to be higher for the CLs than for the NOT CLs, the difference is not statistically significant for the sources considered in this work, and more extended samples are needed to improve the statistics in terms of the X-ray behaviour for CLs. 
 

As a final step we also checked the harder, 2.3--5 keV eROSITA fluxes. Most of the sources have just upper limits in this band, thus we cannot draw further conclusions about the X-ray spectral shape. 

\section{The origin of the NOT CL sources} \label{section:originnotcl}
\begin{figure}
    \includegraphics[width=0.45\textwidth]{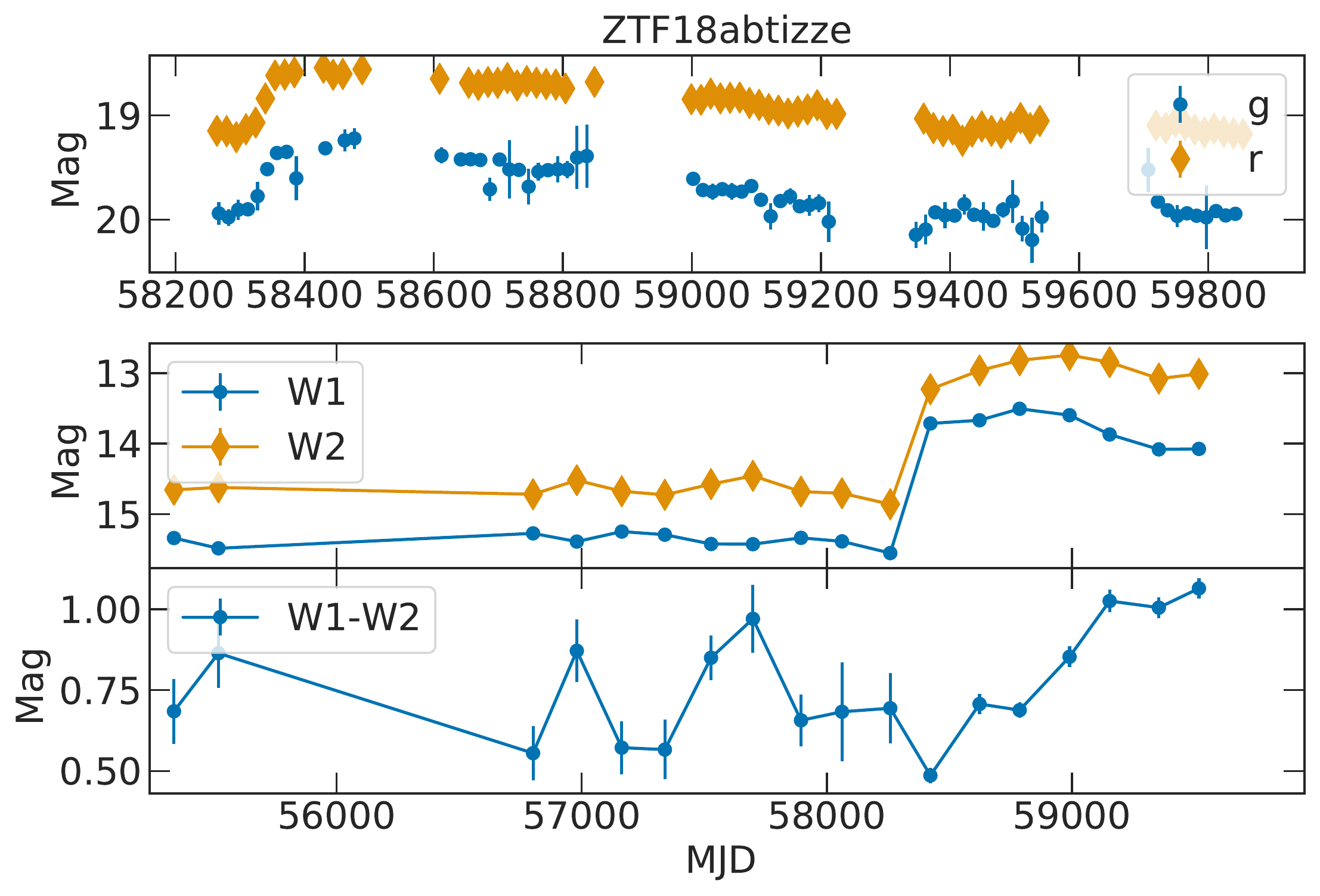}
    
\caption{Optical ZTF forced-photometry light curves and evolution of the \textit{WISE} MIR fluxes and \textit{W1} -- \textit{W2} colour for a TDE candidate belonging to the NOT CL sample. Note the different time-scales of the ZTF and \textit{WISE} data: the optical monitoring starts at the end of the MIR light curves. }
\label{fig:transient}
\end{figure}

The previous section showed that the CL and NOT CL samples are significantly different in terms of their optical, MIR and X-ray flux variability properties, with the CLs showing stronger optical and MIR variability with a tendency to increase their MIR flux and the colour \textit{W1--W2} over time. In order to understand the nature of the NOT CL sources we visually inspected the ZTF forced photometry and alert light curves and the reference images used by ZTF to compute the difference images. In Table~\ref{tab: notcl} we present the characteristics of the variability in individual NOT CL objects, indicating the most probable cause of their variations. As a result, we find two sources that have been most likely misclassified by the LCC due to a small number of alerts (ZTF20aakreaa and ZTF20aaorxzv) and another two sources are possibly false detections due to a bad template image subtraction (ZTF18acgvmzb/ZTF18aclfugf and ZTF18acusqpt/ZTF18adppkkj). Three sources apparently show a transient event (that is, a flare-type variation in an otherwise flat curve) in the optical light curves: two of them resemble SN events (ZTF18aaiwdzt and ZTF18aaqjyon) and one shows a sharp rise in the optical, followed by an MIR echo, which we speculate could be due to a TDE. The occurrence of TDEs in turning-on AGNs is theorised to be more likely than in other galaxies, due to the possibility of `Starfall' \citep{2022mckernan}. This TDE candidate in a Type 2 AGN (shown in Fig.~\ref{fig:transient}) could potentially be happening in a turning-on AGN whose BELs are still too weak to be detected, and merits further study which is beyond the scope of this paper. This discovery highlights the possibility of finding TDE candidates in AGNs, in order to compare their rate of occurrence to TDEs in quiescent galaxies.

\begin{figure}

    \includegraphics[width=0.45\textwidth]{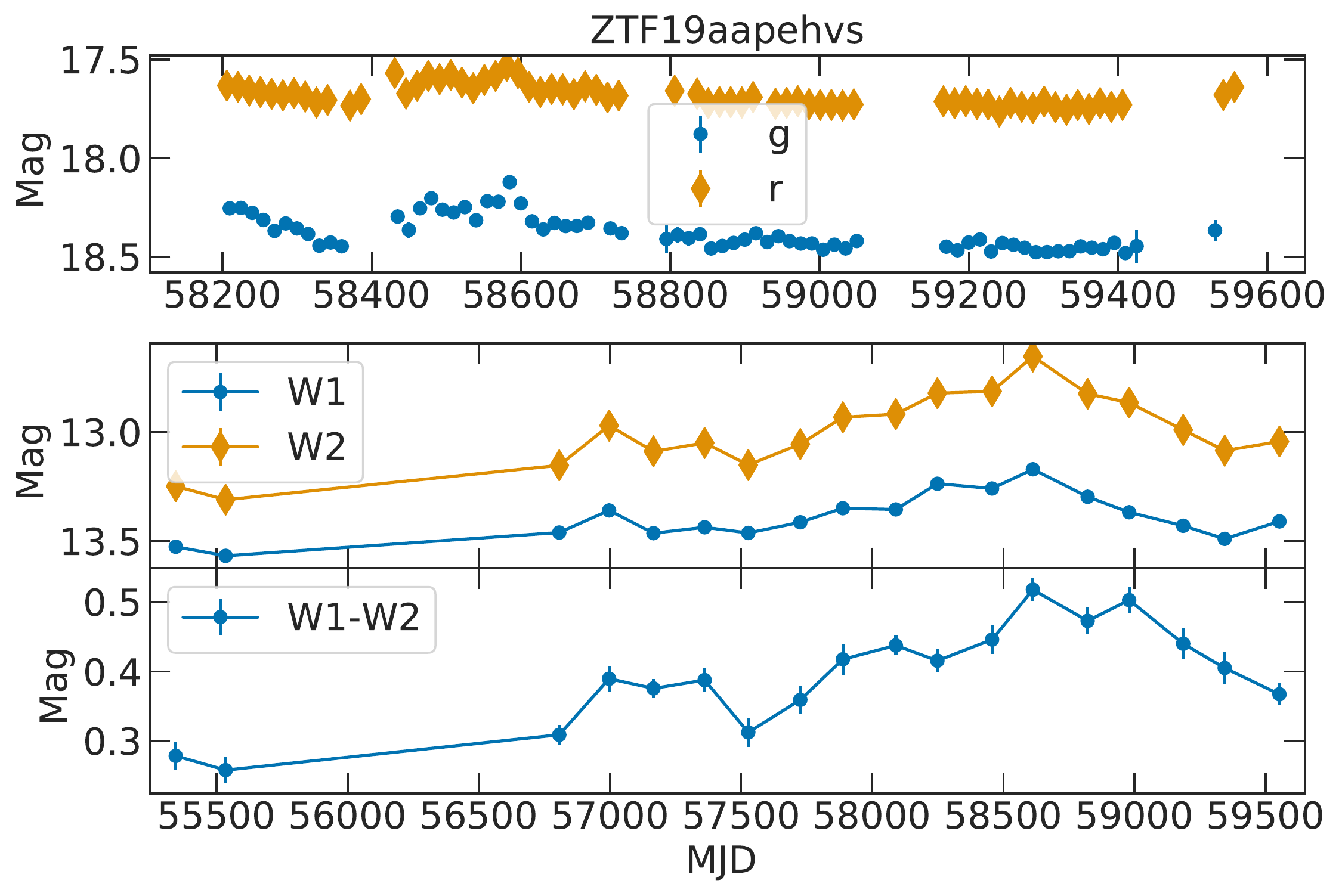}
    \includegraphics[width=0.45\textwidth]{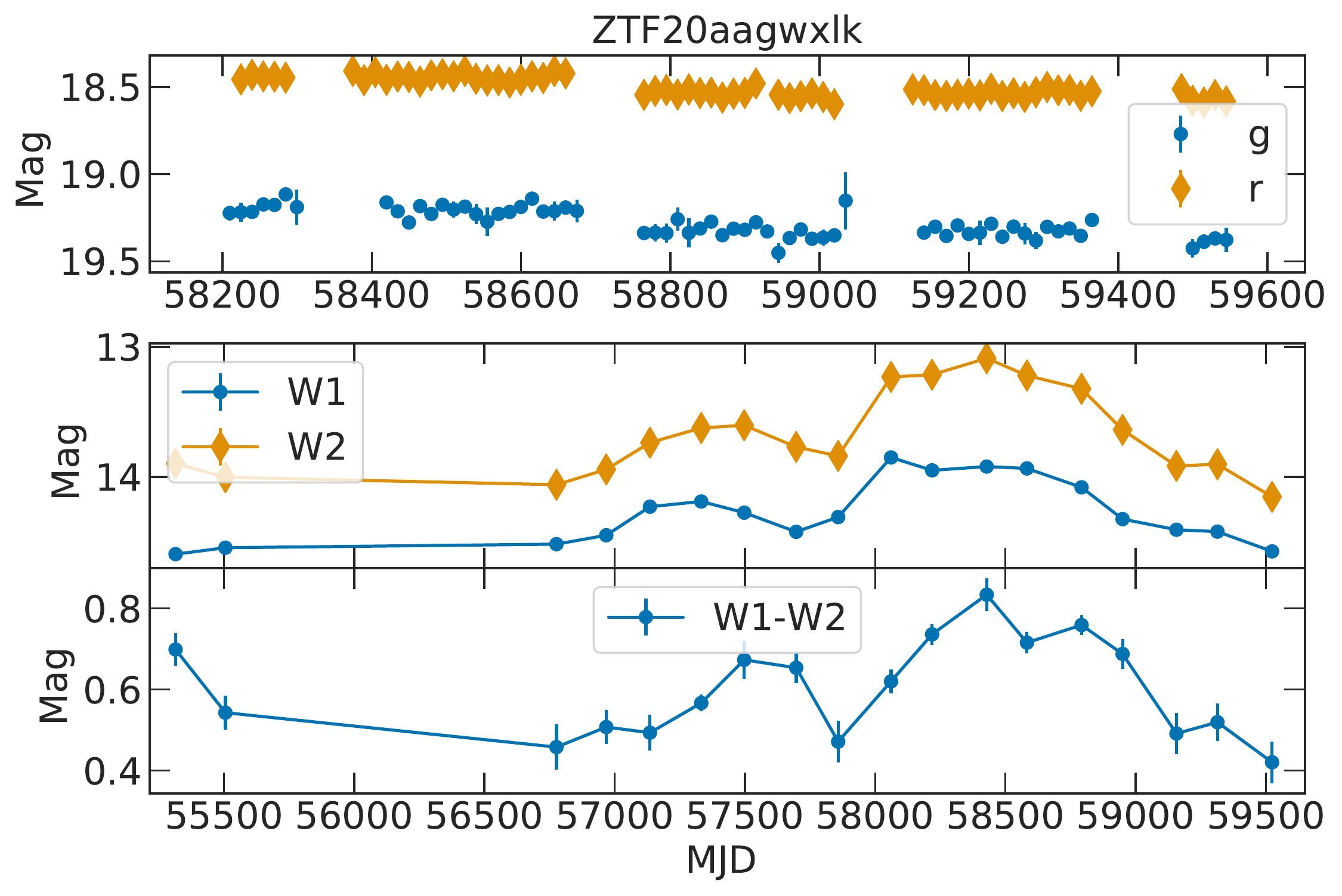}
\caption{Optical ZTF forced photometry light curves and evolution of the \textit{WISE} MIR fluxes and \textit{W1} -- \textit{W2} colour for two NOT CLs sources. Note the different time-scales of the ZTF and \textit{WISE} data: the optical monitoring starts at the end of the MIR light curves, where there is a dimming in the MIR emission. This evolution suggests the sources are now transitioning to a dimmer state.}
\label{fig:notcl}
\end{figure}

Notably, the remaining ten sources show small-amplitude, stochastic optical variations, characteristic of Type 1 AGNs. This is also consistent with their optical spectra, which show weak broad H~$\alpha$ emission lines, indicative of weak Type 1 AGNs. Interestingly, eight of these ten objects show a decrease in their optical flux along with a decrease in their MIR flux and the colour \textit{W1--W2}, which suggests they are now transitioning to a dimmer state. Fig.~\ref{fig:notcl} shows the optical and MIR light curves of two clear examples of this behaviour, ZTF19aapehvs and ZTF20aagwxlk. Generally, the MIR colours from these NOT CL weak Type 1s are galaxy dominated (i.e.,  \textit{W1}--\textit{W2} < 0.5), which suggests their weaker variability and BELs are not due to an orientation effect, but to an intrinsically lower AGN luminosity diluted by the emission of the host galaxy. The remaining two sources, ZTF19aavqrjg and ZTF20abgnlgv, also show Type-1 like variability and stronger broad H~$\alpha$ emission (EW H~$\alpha$ SDSS > 40 \AA), indicative of Type 1 AGNs. 

\label{sec:notcl}
\begin{table*}
	\centering
	    \begin{adjustbox}{max width=\textwidth}
	    \begin{threeparttable}
	\caption{Characterization of the NOT CL sources analysed in this work. \textit{decr.} denotes a decreasing trend in the optical forced-photometry or MIR light curve. \textit{var.} denotes a variable trend. Dashes denote the light curve is fairly flat.  Slashes denote no available eROSITA-DE data. Asteriks denote the two NOT CLs reported in \citetalias{2022lopeznavas}. }
	\label{tab: notcl}

	\begin{tabular}{cccccc} 
ZTF ID & Optical flux &  MIR flux & MIR colour & 0.2--2.3 keV flux & Most likely cause\\
& variability & trend &  \textit{W1}--\textit{W2}   & ($\cdot$ 10$^{-14}$ erg s$^{-1}$ cms$^{-2}$) & \\

\hline
ZTF18aaiescp & yes, decr.  & decr. &  0.5--0.3 &  $/$ & Weak Type 1\\
ZTF18aaiwdzt & transient  &  decr. & 1--0.7  & $/$  & Transient event in the optical light curve\\
ZTF18aaqjyon &  transient & -- & < 0.3 &  $/$ & Transient event in the optical light curve\\
ZTF19aabyvtv/ZTF18aayyapb & yes, decr. & decr.& decr. <0.4 & $/$ & Weak Type 1\\
ZTF18acgvmzb/ZTF18aclfugf & no  & decr. & decr. < 0.5 & < 9.9 & Bogus, bad image subtraction\\
ZTF18achdyst & yes, decr.  & decr. & decr. < 0.5 & 8 $\pm$ 3  & Weak Type 1 \\
ZTF18acusqpt/ZTF18adppkkj & no & var. & < 0.3 & 8 $\pm$ 3  & Bogus, bad image subtraction \\
ZTF19aaixgoj & yes, decr.  & decr. & decr. < 0.5 & <11 & Weak Type 1 \\
ZTF19aapehvs & yes, decr.  & decr. & decr. < 0.5 & $/$ & Weak Type 1\\
ZTF19aavqrjg & yes  & -- & $\sim$0.45 &  6 $\pm$ 3  & Type 1 (BELs)\\
ZTF20aaeutuz & yes, decr.  & decr. & decr. < 0.5 & 13 $\pm$ 5  & Weak Type 1\\
ZTF20aagwxlk & yes, decr.  & decr. & decr. 0.8--0.4 &$/$& Weak Type 1\\
ZTF20aakreaa & no & var. & 0.2--0.6& $/$  & Misclassified, small number of alerts\\
ZTF20aaorxzv & no & -- & $\sim$ 0.2 & < 9.5  & Misclassified, small number of alerts\\
ZTF20abgnlgv & yes & var. & 0.4--0.6 & $/$ & Type 1 (BELs)\\
ZTF18abtizze * & yes & var. & 0.4--1 & $/$ & Possible TDE \\
ZTF19aaxdiui * & yes, decr. & decr. & decr. 0.6--0.2& $/$ & Weak Type 1\\

    \hline
	\end{tabular}
	\end{threeparttable}
	\end{adjustbox}
\end{table*}

\section{Discussion}
\subsection{Improvement of the CL selection method through ALeRCE} \label{improvement}

One of the main aspects of the selection of CL candidates through ALeRCE relies on the correct classification of their alert light curve variability by the LCC. As mentioned in Section \ref{section:originnotcl}, there are seven out of 17 NOT CLs that have been misclassified by the LCC due to a bad subtraction of the ZTF images used to compute the alerts, a small number of data points and/or transient events in the alert light curves. On the other hand, we also found that ten ($\sim$60 per cent) NOT CLs have been correctly classified by the LCC as Type 1 AGN and are spectroscopically consistent with being weak Type 1 AGNs whose optical and MIR properties are dominated by the host galaxy contribution.

It is possible to increase the completeness of the CL candidate list as well as the purity by applying two additional criteria on the ALeRCE features simultaneously. The best combinations in the \textit{g} band, with a loss of 0--5 per cent of CLs and removal of 50 per cent of NOT CLs, were found to be all the possible pairs between GP$\_$DRW$\_$sigma, GP$\_$DRW$\_$tau and delta$\_$mag$\_$fid, and the combinations of iqr with MHPS$\_$low or delta$\_$mag$\_$fid. We note that our samples are too small to judge the universality of this additional cleaning with bootstrap methods. However, similar cleaning can be devised in the future for ZTF forced photometry and/or the classifications of the ZTF data release light curves \citep{2023sanchez}, which would help to improve the selection of CL candidates.

\begin{table}
	\centering
	    \begin{adjustbox}{max width=\textwidth}
	    \begin{threeparttable}
	\caption{Variability and colour features in the \textit{g} band that can be used to clean the CL candidate list. The limiting lower values correspond to the 16th percentile obtained for the CL sample. The third column indicates the percentage of NOT CL sources that can be discarded by limiting each feature to its limiting value.}
	\label{tab:cleaning}

	\begin{tabular}{ccc} %
\textbf{Feature}&\textbf{Limiting value}&\textbf{$\%$ cleaning}\\
\hline
positive$\_$fraction & 0 & 56 \\
SF$\_$ML$\_$amplitude & $-$0.2 & 56\\
SPM$\_$tau$\_$rise & 44 & 50 \\
GP$\_$DRW$\_$sigma & 4$\cdot 10^{-5}$& 56 \\
GP$\_$DRW$\_$tau  & 9 & 56\\
IAR$\_$phi & 0.97 & 44\\
MHPS$\_$low & 0.012 & 63\\

delta$\_$mag$\_$fid & 0.17 & 56\\
r–\textit{W3} & 6.7 & 6 \\
g$\_$r$\_$mean$\_$corr & 0.60 & 18 \\
g$\_$r$\_$max & $-$0.10 & 12\\

n$\_$pos & 0 & 56\\
iqr & 19.7 & 63\\

\hline
    \end{tabular}
    \end{threeparttable}
    \end{adjustbox}
\end{table}

\subsection{Characterizing the turning-on  CL properties}
Apart from the ALeRCE LCC features, we have analysed other multi-wavelength properties of the sample to investigate what can increase the likelihood of finding a turning-on transition.  
\subsubsection{BPT diagrams}
 While most of the reported CLs in the literature are Seyferts, some of them have been found to lie on the borderline between a LINER and Seyfert classification, with extreme order-of-magnitude changes in continuum and emission-line flux compared to less dramatic CLs occurring in Seyferts \citep{frederick2019}.
 
 Here, we tested whether the CL transitions could be related to the emission-line properties of the sources by computing the line-ratio diagnostic diagrams involving the line ratios [O \textsc{iii}]/H~$\beta$, [N \textsc{ii}]/H~$\alpha$, [O \textsc{i}]/H~$\alpha$, and [S \textsc{ii}]/H~$\alpha$ \citep{1981BPT,2006kewley}. By selection, all our CL candidates are consistent with a Seyfert classification, and the BPT criteria indicate that the line ratios from both their archival and new optical spectra distribute similarly, regardless of whether they experience a CL transition or not. 

\subsubsection{Forced photometry variability}

 Due to a higher number of data points, analyzing the complete light curves (instead of the alert curves) gives us a much more reliable determination of the variability of the sources. In general, we find a stronger variability typical of luminous Type 1 AGNs in the CL sample (during or post transition), which indicates that most of the candidates with non-variable BELs (i.e., the NOT CL sample) belong to a different population. Recently, ZTF light-curve variability analysis has been found to be a powerful tool to seek new CL candidates. 
In \citet{2023lopeznavas}, by analysing ZTF forced photometry light curves of $\sim$ 15000 Type 2 AGNs, we were able to distinguish between weak Type 1 and Type 2 sources and select CL candidates that have also been found through ALeRCE in the work presented here. \citet{sanchez2021_anomaly} used light curves from ZTF data release 5 (ZTF DR5) instead, and applied deep learning techniques to find anomalous behaviour in AGNs, leading to the identification of 75 promising CS candidates. 



\subsubsection{MIR variability}

The study of the photometric variability in the MIR band can help us understand the physical mechanisms that may be involved in CL events. This emission is believed to come from dust thermally heated by the AGN, which emits relatively unimpeded by dust extinction. Indeed, simple  \textit{W1}--\textit{W2}  (i.e., $[3.4]-[4.6] \mu$m) colour cuts with \textit{WISE} data are able to reliably differentiate AGNs from stars and galaxies \citep[e.g.  \textit{W1}--\textit{W2}  $\geq$ 0.8, ][]{2012Stern}. Along these lines, an important feature in our analysis is a$_{12}$, which indicates the linear trend of the  \textit{W1--W2} colour. Our CL sample shows a general upward trend during the last $\sim$11 years, indicating a redder colour in the MIR band. This behavior is consistent with a turning-on transition, where the sources change from galaxy-like (i.e.,  \textit{W1}--\textit{W2}  <0.5) to AGN-like (i.e.,  \textit{W1}--\textit{W2}  > 0.8) MIR colour, as it has been found in previous CL studies \citep{2017sheng,yang2018,2020sheng,2022lyu}. This phenomenon can be explained by an increase of the brightness of the AGN, which illuminates the torus and produces a stronger contribution of hot dust in the \textit{W1} and \textit{W2} bands, with a larger effect on the latter. This scenario is supported by the observed \textit{W1} and \textit{W2} fluxes in our CL sample, which also show an upward trend given by the slopes a$_1$ and a$_2$. 

To estimate the variability of the CLs in comparison with the NOT CL sample, we computed the parameters $\langle\sigma_{m1}\rangle$ and $\langle\sigma_{m2}\rangle$ (indicative of the intrinsic variability) and the maximum variations of  \textit{W1}  and  \textit{W2}  ($\langle\Delta$\textit{W1}$\rangle$ and $\langle\Delta$\textit{W2} $\rangle$) as in \citet{2022lyu}. Those authors present a study of the variability of the \textit{W1}  and \textit{W2} light curves of a population of CLs in comparison with low-luminosity AGN and QSO samples. They also use the AllWISE multi-epoch photometry and NEOWISE single exposure (L1b) source table data, though we average the data every six months. Our results, presented in Table \ref{tab:mir}, are consistent with their conclusions. On the one hand, we find that the CL population has a higher $\langle\sigma_{m1}\rangle$ and $\langle\sigma_{m2}\rangle$ variability than the NOT CL population. On the other hand, the CL sample exhibits maximum variations of \textit{W1} and  \textit{W2} ($\langle\Delta$\textit{W1}$\rangle$ and $\langle\Delta$\textit{W2}$\rangle$) greater than 0.3 mag. In the scenario of variable obscuration, \citet{yang2018} estimate that a variation in the  \textit{W1}  band due to dust extinction would yield a factor of $\sim$21 change in the \textit{g} band magnitude, assuming the extinction curve in the optical and MIR and considering micrometer-sized grains \citep{2015wang}. A variation of 0.3 mag in the  \textit{W1}  band would lead to $\sim$6.3 mag difference in the \textit{g} band, which is not consistent with the observed properties of CLs.  Therefore, the variability and general increase in the MIR flux and the colour \textit{W1--W2} of the CL sample, over the last 11 years, are most likely due to an intrinsic brightening of the AGN and cannot be solely explained by the motion of absorbing clouds within our line of sight.

\subsubsection{X-rays}

The CL sample shows higher X-ray fluxes and higher X-ray to \textit{g}-band flux ratios than the NOT CL sample in the recent  eROSITA scans. We note that the AGN continuum in the optical spectra of these sources is sub-dominant or undetectable under the starlight contribution, so a significant fraction of the \textit{g}-band flux can be associated with the host galaxy. For a given host-galaxy flux, the CL sources are X-ray brighter, as expected if their AGNs are more dominant than in the NOT CL sample.

There is a relatively small number of CL sources with contemporaneous X-ray data in the literature, so  it is still unclear whether these events are always accompanied by a clear change in the X-ray band. We note that we are talking about optical CLs, not X-ray CLs where the changes are due to obscuration. In the well studied case of the CL Mrk 1018, a strong soft X-ray excess was found when the source was in the bright state, which dropped by much more than the hard X-rays when the source transitioned to the dimmest state \citep{2016mcelroy, 2016husemann}. In this case, the changes in the BELs were associated with an increase/decrease of ionizing photons from the soft X-ray excess \citep{2018noda}. Additionally, \citet{2022temple} found that five out of nine CL events with \textit{Swift}-BAT data available could be associated with significant X-ray changes in the hard 14–195 keV band. The hard band is less affected by obscuration and thus rules out a varying column density as the origin of the flux changes. 
In this work, we find further evidence that at least some of the CL transitions could be associated with significant X-ray flux changes. In particular, four out of the 11 CLs with eROSITA data have experienced an X-ray flux rise with respect to archival \textit{ROSAT} data taken 30 years earlier. These sources also show variable flux within the eROSITA 2019--2022 light curves, with the X-ray flux being four, eight and up to 15 times higher than the minimum value (see Fig.~\ref{fig: xraylc}). However, due to the small number of data points and the intrinsic X-ray variability in unobscured AGNs, we are unable to discard the possibility that these changes are purely stochastic and not related to the CL event. 

The CL AGN that has been most extensively observed in X-ray monitoring is 1ES 1927+654. This source was found to show dramatic changes in its X-ray spectral shape, showing the disappearance and re-appearance of the X-ray corona after the CL event \citep{trakhtenbrot2019, ricci2020, 2021ricci}. The authors suggest that this particular phenomenon was caused by a TDE in an AGN. This strengthens  the importance of multi-wavelength campaigns studying CLs, which enable constraints on the evolution of the AGN components during the transitions.



\subsection{Promising CL candidates selected through multi-wavelength flux variability}
The multi-wavelength photometric analysis performed in this work provides a powerful tool to identify turning-on CL events, allowing us to select the most promising CL events from the candidate list we have not re-observed. In Table \ref{tab: promising} we show the sources from our CL candidates that are currently varying in the optical according to the refined ZTF feature criteria (all the possible pairs of best combinations mentioned in Section  \ref{improvement}), and that are experiencing a MIR flux and colour \textit{W1--W2} increase over time. For the sources with eROSITA-DE data, their X-ray fluxes were below the archival limits, so we did not use X-ray data as a selection criterion. In Fig.~\ref{fig: promising} we present the optical and MIR light curves for these sources. Some of them were not selected to be re-observed because they already showed obvious BELs in their archival optical spectra, such as ZTF18aaodaie and ZTF21abmjobi. However, according to their increase in the MIR fluxes and colour, it is likely that these sources now present stronger BELs and are therefore worth re-observing. We strongly encourage follow-up spectroscopy to confirm the CL behaviour in the promising candidates presented here, which will improve the statistics and refine further the selection criteria of CL candidates.
We note that  the cleaning criteria in the optical and the evolution of the MIR flux and color used to select the most promising candidates are fairly independent for the redshifts considered in this work, from $z=0 $ to $z=0.4$.

\begin{table*}
	\centering
	    \begin{adjustbox}{max width=\textwidth}
	    \begin{threeparttable}
	\caption{Potential CL sources from our list of CL candidates identified through their current optical variability and the criteria mentioned in the comments.}
	\label{tab: promising}

	\begin{tabular}{cccccccc} %
Name & Ra & Dec & z & MJD & fiberid & plate & Comments  \\
\hline

ZTF18aalmtum & 136.365522 & 17.928647 & 0.40 & 56246 & 330 & 5769 & Increasing MIR fluxes and  \textit{W1}--\textit{W2}.\\
ZTF18aaodaie/ZTF21ackxdiy & 191.572165 & 28.342741 & 0.10 & 54205 & 442 & 2238 & Increasing MIR fluxes and  \textit{W1}--\textit{W2}.\\ 
ZTF18aaoudgg & 221.975974 & 28.556697 & 0.16 & 53764 & 134 & 2141 & Increasing MIR fluxes and  \textit{W1}--\textit{W2}. \\
ZTF18acakoya/ZTF18abtndvk & 43.709715 & -2.797898 & 0.12 & 56978 & 181 & 7823 & Increasing MIR fluxes and  \textit{W1}--\textit{W2}.\\
ZTF19adcfhxp & 143.823381 & 34.320597 & 0.45 & 56336 & 727 & 5805 & Increasing MIR fluxes and  \textit{W1}--\textit{W2}.\\
ZTF20acpcfgt & 339.995984 & 0.860685 & 0.38 & 52201 & 620 & 674 & Increasing MIR and optical fluxes and  \textit{W1}--\textit{W2}.\\
ZTF21abjprnu & 230.292534 & 20.544009 & 0.14 & 54328 & 421 & 2159 & Increasing MIR and optical fluxes. Decreasing  \textit{W1}--\textit{W2}, but still \textit{W1}--\textit{W2}  > 0.7\\
ZTF21abdvcpz & 335.843667 & 0.771973 & 0.22 & 52140 & 454 & 375 & Increasing MIR fluxes and  \textit{W1}--\textit{W2}.\\
ZTF21abmjobi & 358.363382 & 0.123463 & 0.17 & 52523 & 520 & 684 &  Increasing MIR and optical fluxes and  \textit{W1}--\textit{W2}.\\

\hline
    \end{tabular}
    \end{threeparttable}
    \end{adjustbox}
\end{table*}

\subsection{Key questions about the CL phenomena}

\subsubsection{Frequency and time-scales}
CL events have been found to occur on time-scales as short as one or two months \citep{trakhtenbrot2019, 2022zeltyn}. However, most research to date uses archival spectral and/or photometric data to look for CL transitions, making it difficult to obtain a proper estimate of the typical CL time-scales and thus understand the drivers of such extreme variations \citep{2018stern}. The time difference between the first and last spectral epochs for our CL sample lies between 10 and 20 years (in the observed frame), which must be taken as an upper limit on the time-scale for these transitions. Given their low incidence rate (discussed below) it is difficult to find a large amount of CL events as they take place. However, much larger and/or intensively sampled spectroscopic campaigns, such as SDSS-V \citep{2017Kollmeier} and 4MOST \citep{20194most} will provide better constraints on the typical CL time-scales, potentially finding shorter (weeks and even days) CL events.

Another key question we can address with our results is the occurrence rate of turning-on transitions, following the discussion in \citetalias{2022lopeznavas}. There are 30,333 AGNs classified as Type 2 (N and K types in MILLIQUAS, cleaned for weak Type 1s and LINERS) that can be detected by ZTF (Dec $>-28 ^\cdot$ and $r<21$ mag). From them, 178 have alerts and are classified as AGN or QSO by the LCC. The 50 per cent confirmation rate found in this work implies that 0.3 per cent of transitions can be detected by this method over an average timespan of 15 years. Taking into account that just 10 per cent of Type 1 AGNs are variable enough to generate alerts in ZTF, we estimate a lower limit of 3 per cent of turning-on events every 15 years (i.e., 0.2 per cent per year). These values are consistent with the results from \citet{2022hon}, who finds a minimum turn-on CL AGN rate of 3 per cent every 15 years, and from \citet{2022temple}, who reports a CL rate of 0.7--6.2 per cent on 10--25 year time-scales (including both turning-on and turning-off events).
As a difference with respect to the present work, \citetalias{2022lopeznavas} selected the most promising candidates via visual inspection of the light curves and archival spectra, reducing the number of candidates and therefore the ratio between CL candidates to the number of objects in the parent sample. Therefore, even though the confirmation rate was higher than in this manuscript, they obtained a lower estimate of the total rate of change (0.12 per cent per year). In the present work we did not further restrict the candidate list to the most promising candidates so the fraction of the parent sample that was considered as a candidate is larger. Even if our success rate is slightly smaller, the product of both factors leads to a somewhat larger total rate of change. We note that in both cases we are estimating lower limits to the rate of change.

\subsubsection{Physical origin}
In principle, several mechanisms could lead to variations in the BELs, including variable obscuration and changes in the accretion rate. Here, we find that the MIR emission, which is dominated by the dust response (i.e., from the torus) to the UV–optical variations of the central engine, tends to increase in the CL sources. Since the MIR emitting region is too large to be obscured/unobscured on time-scales of <20 years, and the emission is much less affected by dust obscuration, we infer that the changes in the MIR waveband and by extension the CL transitions are due to changes in the accretion rate. 

To explain the changes in the accretion flow at these time-scales, the proposed scenarios include instabilities in the accretion disc and major disc perturbations such as those caused by TDEs. In general, the light curves of our CL sources do not follow the power law decay t$^{-5/3}$ from the peak brightness as traditionally expected for TDEs \citep{1990rees}. On the other hand, the values of $\lambda_{\rm   \rm Edd}$ obtained for the CL sample in their bright state are similar to the results for CL AGNs and QSOs reported in recent works  \citep[$-2 \lessapprox $ log$\lambda_{\rm   \rm Edd} \lessapprox -1$,][]{macleod2019, frederick2019,graham2020, 2022temple}, and is consistent with  attributing  the  changes in accretion rate occurring preferentially in lower activity systems. Interestingly, our sources have an Eddington ratio (post-CL) in the range  1--5 per cent $L_{\rm Edd}$, which is in agreement with the expectations from a hydrogen ionization disc instability \citep{2018noda,2019ruan, 2023sniegowska}. In this scenario, by analogy to the spectral transitions in black hole X-ray binaries, the changes in the structure of the inner accretion disc occur around a critical value of  $\lambda_{\rm   \rm Edd} \sim$ 0.02, which is in agreement with the most recent CL studies \citep{graham2020, 2021guolo, 2022temple}. Furthermore, our CL sources show a redder-when-brighter tendency in the MIR, with AGN-like MIR colours (i.e.  \textit{W1}--\textit{W2}  > 0.5) when they enter the bright state, which also has been found in other CL sources \citep{yang2018,2022lyu} and supports the accretion state transition. 

In this scenario, either the BELs appear due to the increase of ionizing photons that excite the BLR \citep{lamassa2015}, or the BLR itself re-appears according to the expectations in disc-wind BLR models \citep{2000Nicastro,2009elitzur}. These models predict an evolutionary sequence for the BLR depending on the accretion rate, leading to different intermediate type spectral transitions and the BLR disappearance at very low luminosities \citep[Lbol $\lessapprox 5 \cdot 10^{39}M^{2/3}_{7}$ erg s$^{-1}$, where $M_{7}=M/10^{7} $M$_\odot$, ][]{2009elitzur, 2014elitzur}. According to the Eddington ratio estimates, none of our sources fell close to these limits, which suggests the BLR already existed in these sources but was too weakly ionised to produce detectable broad lines.  On the other hand, the residual spectra in Fig. \ref{fig: lc and spec} show that the continuum of the CLs currently looks either flat or blue. This `bluer when brighter' effect, although clearly real in quasars, could at least partially be due to differences in the relative contribution of the host galaxy as these lower-luminosity AGNs change their intrinsic flux.

\section{Summary and conclusions}
This paper is the continuation of work introduced in \citetalias{2022lopeznavas}, where we present a method to search for turning-on CL candidates. The selection method consists of searching for current Type-1 AGN variability in a sample of spectrally classified Type 2 AGNs, using classifications given by a random-forest based light curve classifier, the ALeRCE LCC \citep{sanchez2021}. In order to refine the selection method, we obtained second epoch spectra for 36 of our 61 CL candidates, six of which were reported in \citetalias{2022lopeznavas}, which allows the confirmation of the CL objects by quantifying the change of the BELs in comparison with the archival SDSS spectra taken 15--20 years earlier. As a result, we find 19 ($\sim$ 50 per cent) turning-on CL confirmations (the CL sample), and 17 sources without significant changes in their BELs (the NOT CL sample).

We have analysed the multi-wavelength properties of the CL sample in comparison with the NOT CL sample to investigate what would increase the likelihood of finding a turning-on transition and understand its origin. Firstly, we performed a variability analysis of the alert light curves from ZTF, finding several variability features that are distinct between the samples and that can be applied to select the most promising CL sources. We also find that the turning-on transitions are characterised by an increase in the \textit{WISE}-MIR brightness and the MIR (\textit{W1}--\textit{W2}) colour, where the stronger H~$\alpha$ emission corresponds to an AGN-dominated MIR colour (\textit{W1}--\textit{W2} > 0.5 mag). The current Eddington ratio estimations for the CLs are lower than the overall Type 1 population, falling between one and five per cent $L_{\rm Edd}$. In the X-ray band, we find that the CLs tend to be flux brighter than the candidates that have not transitioned, and for four CLs we observe a significant flux increase during the 2019--2022 X-ray monitoring. These results are in agreement with previous CL/CS works, and with the expectations from an accretion-state transition as the origin of these phenomena. 

We also analyse the nature of the NOT CL sources according to their optical and MIR variability. For seven out of the 17 objects, the Type 2 sources were misclassified by the LCC due to a bad subtraction of the images, a small number of data points and/or transient events in the light curves such as SNe and one case of a likely TDE. Interestingly, we also find that ten sources are consistent with a Type 1 classification, where the optical and MIR emission is dominated by the host galaxy. This translates into lower amplitude variations in the optical and MIR wavebands, weaker BELs and galaxy-dominated MIR colours (i.e., \textit{W1}--\textit{W2} <0.5). Incidentally, we also find that seven of the NOT CL sources are currently decreasing their optical and MIR fluxes, suggesting they are currently transitioning to a dimmer state. The multi-wavelength differences between the CL and NOT CL sources allow us to select the most promising CL candidates from our list without spectroscopic follow-up, leading to nine sources that are worth re-observing.

The use of machine learning algorithms on complete optical light curves from the ZTF or the upcoming LSST can be combined with MIR data to unequivocally identify CLs, improve the statistics and ultimately understand the underlying physics of these phenomena.

\section*{Acknowledgements}

ELN and SB acknowledge support from Agencia Nacional de Investigación y Desarrollo (ANID) / Programa de Becas/ Doctorado Nacional 21200718 and 21212344. ELN acknowledges the California Institute of Technology and the European Southern Observatory for their hospitality. PA, ELN, MLMA and PL acknowledge financial support from Millenium Nucleus NCN$19\_058$ (TITANs). PA acknowledges financial support from the Max Planck Society through a Partener Group with the between MPA and the University of Valparaíso. LHG acknowledges funds by ANID – Millennium Science Initiative Program – ICN12$\_$009 awarded to the Millennium Institute of Astrophysics (MAS). PL acknowledges partial support from FONDECYT through grant Nº 1201748. PSS acknowledges funds by ANID grant FONDECYT Postdoctorado Nº 3200250. MJG acknowledges partial support from the NSF grant AST-2108402. DT acknowledges support by DLR grant FKZ 50 OR 2203. DH acknowledges support from DLR grant FKZ 50 OR 2003. MK acknowledges support from DFG grant number KR3338/4-1. Based on observations collected at the Samuel Oschin Telescope 48-inch and the 60-inch Telescope at the Palomar Observatory as part of the Zwicky Transient Facility project. The ZTF forced-photometry service was funded under the Heising-Simons Foundation grant 12540303 (PI: Graham). This publication makes use of data products from the Wide-field Infrared Survey Explorer, which is a joint project of the University of California, Los Angeles, and the Jet Propulsion Laboratory/California Institute of Technology, funded by the National Aeronautics and Space Administration. This publication also makes use of data products from NEOWISE, which is a project of the Jet Propulsion Laboratory/California Institute of Technology, funded by the Planetary Science Division of the National Aeronautics and Space Administration. This work is based on data from eROSITA, the soft X-ray instrument
aboard SRG, a joint Russian-German science mission supported by the
Russian Space Agency (Roskosmos), in the interests of the Russian
Academy of Sciences represented by its Space Research Institute (IKI),
and the Deutsches Zentrum f\"ur Luft- und Raumfahrt (DLR). The SRG
spacecraft was built by Lavochkin Association (NPOL) and its
subcontractors, and is operated by NPOL with support from the Max
Planck Institute for Extraterrestrial Physics (MPE). The development
and construction of the eROSITA X-ray instrument was led by MPE, with
contributions from the Dr. Karl Remeis Observatory Bamberg \& ECAP
(FAU Erlangen-Nuernberg), the University of Hamburg Observatory, the
Leibniz Institute for Astrophysics Potsdam (AIP), and the Institute
for Astronomy and Astrophysics of the University of T\"ubingen, with
the support of DLR and the Max Planck Society. The Argelander
Institute for Astronomy of the University of Bonn and the Ludwig
Maximilians Universit\"at Munich also participated in the science
preparation for eROSITA.

The eROSITA data shown here were processed using the eSASS/NRTA software system developed by the German eROSITA consortium.

\section*{Data Availability}

The SDSS data underlying this article were accessed from SDSS DR17 (\url{http://skyserver.sdss.org/dr17}). The second epoch spectra data will be shared on reasonable request to the corresponding author. The MIR data are publicly available at \url{https://irsa.ipac.caltech.edu/Missions/wise.html}. The alert ZTF light curves, together with the LCC classifications, can be downloaded at \url{https://alerce.online}. The forced-photometry ligth curves can be requested  via the ZTF photometry service. The eROSITA data underlying this article were provided by the eROSITA-DE collaboration by permission, and will be shared on request to the corresponding author with permission of the eROSITA-DE collaboration. eRASS-1 will be public from Fall 2023. 



\bibliographystyle{mnras}
\bibliography{example} 



\appendix

\section{Light curves and spectra from the CL sample}
\begin{figure*}
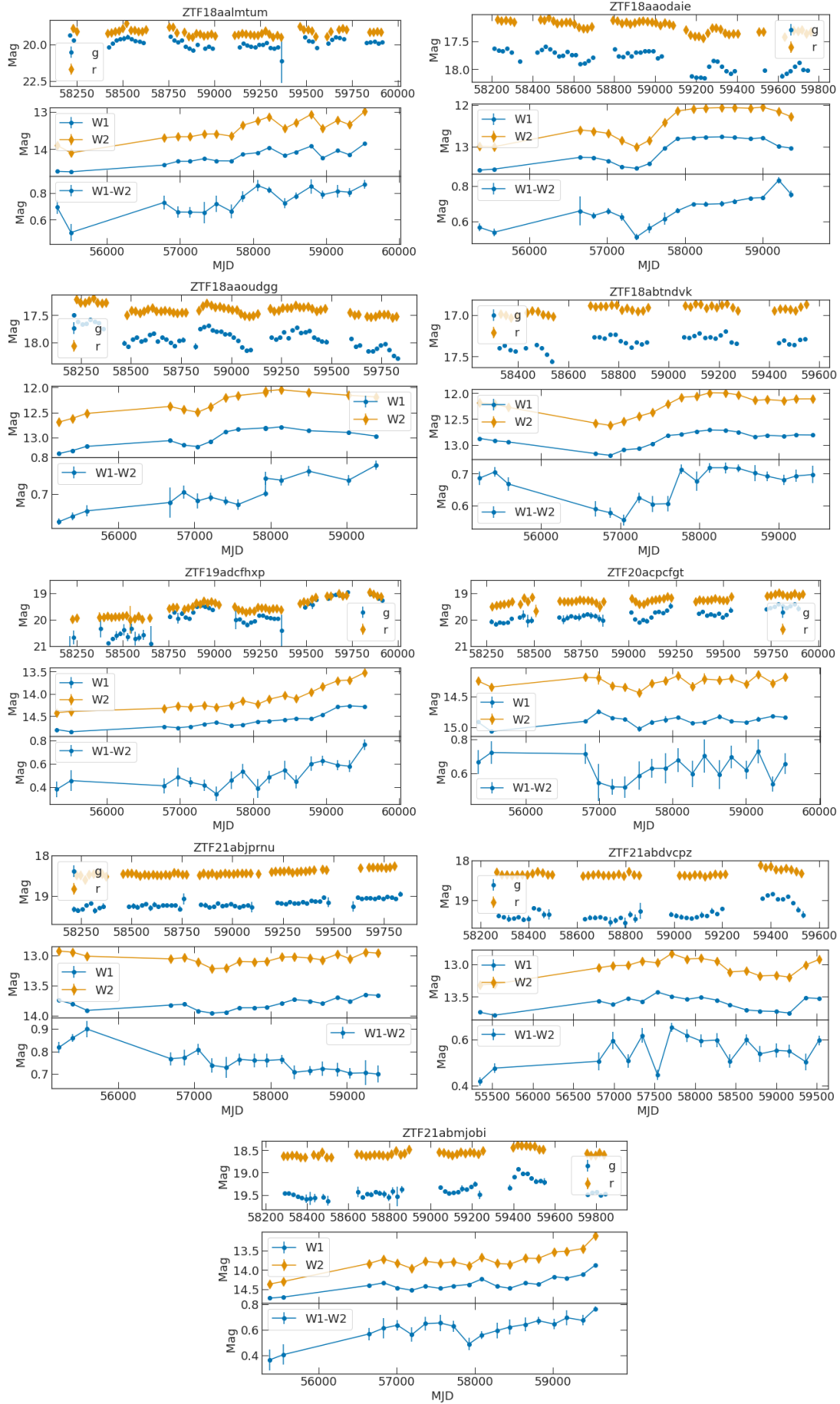

        \centering
            \begin{subfigure}{0.4\textwidth} 
            \includegraphics[width=\textwidth]{plots/ZTF18aalmtum.pdf}
        \end{subfigure}    
        \begin{subfigure}{0.4\textwidth} 
            \includegraphics[width=\textwidth]{plots/ZTF18aaodaie.pdf}
        \end{subfigure}

                    \begin{subfigure}{0.4\textwidth} 
            \includegraphics[width=\textwidth]{plots/ZTF18aaoudgg.pdf}
        \end{subfigure}    
        \begin{subfigure}{0.4\textwidth} 
            \includegraphics[width=\textwidth]{plots/ZTF18abtndvk.pdf}
        \end{subfigure}    

                    \begin{subfigure}{0.4\textwidth} 
            \includegraphics[width=\textwidth]{plots/ZTF19adcfhxp.pdf}
        \end{subfigure}    
        \begin{subfigure}{0.4\textwidth} 
            \includegraphics[width=\textwidth]{plots/ZTF20acpcfgt.pdf}
        \end{subfigure}   

                            \begin{subfigure}{0.4\textwidth} 
            \includegraphics[width=\textwidth]{plots/ZTF21abjprnu.pdf}
        \end{subfigure}    
        \begin{subfigure}{0.4\textwidth} 
            \includegraphics[width=\textwidth]{plots/ZTF21abdvcpz.pdf}
        \end{subfigure}   
\begin{subfigure}{0.4\textwidth} 
            \includegraphics[width=\textwidth]{plots/ZTF21abmjobi.pdf}
        \end{subfigure}    
        
     \caption{ZTF optical and \textit{WISE}-MIR light curves from the most promising CL candidates.} 
\label{fig: promising}

\end{figure*}

Table ~\ref{tab: new spectra} indicates the dates and instruments used for the second epoch spectra of all the CL candidates observed in this work.
Fig.~\ref{fig: lc and spec} shows the ZTF optical and \textit{WISE}-MIR light curves (left) and optical spectra (right) for the 15 CL sources identified in this paper. The optical spectra are scaled to the flux of [S \textsc{ii}] in the earliest spectra and smoothed with a 10 \AA\  box filter. The lower plots in the right side show the difference between the new and the old spectra. In some cases, the new spectra were taken with a blue and a red arms, leading to very noisy regions that have been deleted. The other four CL sources considered in this work are reported in \citet{2022lopeznavas}. All the SDSS spectra analysed in this work were taken before the beginning of the \textit{WISE}-MIR light curves, except for the case of ZTF20aagyaug (MJD SDSS=55673). The second epoch spectra were taken between MJD 59600 and 59700, at the end of the optical and MID-IR light curves.
Fig.~\ref{fig: promising} shows the ZTF optical and \textit{WISE}-MIR light curves for the most promising CL candidates selected according to their optical and MIR photometric variability.

\begin{table}
	\centering
	    \begin{adjustbox}{max width=\textwidth}
	    \begin{threeparttable}
	\caption{Dates and instruments of the second epoch spectra for the CL candidates observed in this work. Sources identified as CL are shown in bold. }
	\label{tab: new spectra}

	\begin{tabular}{ccccc} 
ZTF ID & RA & DEC & MJD & Instrument\\
& $\deg $& $\deg$ & & \\
\hline
ZTF18aaiescp & 207.21292 & 57.646792 & 59603 & DBSP(P200)\\
ZTF18aaiwdzt & 199.48361 & 49.258651 & 59603 & DBSP(P200) \\
\textbf{ZTF18aajywbu} & 205.45327 & 37.013091 & 59604 & LRIS(Keck I) \\
\textbf{ZTF18aaqftos} & 180.95505 & 60.888181 & 59696 & DBSP(P200)\\
ZTF18aaqjyon & 180.42264 & 38.47264 & 59696 & DBSP(P200) \\
\textbf{ZTF18aasudup} & 170.03619 & 34.312731 & 59603 & DBSP(P200)\\
\textbf{ZTF18aavxbec} & 243.08151 & 46.495172 & 59696 & DBSP(P200) \\
\textbf{ZTF18aawoghx} & 156.10498 & 37.650863 & 59603 & DBSP(P200)\\
\textbf{ZTF18aawwcaa} & 128.10120 & 35.859979 & 59603 & DBSP(P200) \\
ZTF19aabyvtv & 197.87773 & 31.866893 & 59603 & DBSP(P200)\\
\textbf{ZTF18acbzrll} & 124.82294 & 30.32660 & 59696 & DBSP(P200) \\
ZTF18acgvmzb & 148.96896 & 35.965616 & 59616 & DBSP(P200) \\
ZTF18achdyst & 157.39925 & 24.777606 & 59616 & DBSP(P200) \\
ZTF18acusqpt & 177.91898 & 12.036714 & 59603 & DBSP(P200)\\
\textbf{ZTF19aafcyzr} & 125.71008 & 15.673859 & 59616 & DBSP(P200) \\
ZTF19aaixgoj & 146.80538 & 12.205624 & 59603 & DBSP(P200) \\
\textbf{ZTF19aaoyjoh} & 180.18956 & 14.967685  & 59603 & DBSP(P200) \\
ZTF19aapehvs & 199.74519 & 57.501847 & 59696 & DBSP(P200)\\
ZTF19aavqrjg & 181.24583 & 15.58718  & 59603 & DBSP(P200)\\
\textbf{ZTF19aavyjdn} & 202.39941 & $-$1.509453 & 59603 & DBSP(P200) \\
ZTF20aaeutuz & 164.81581 & 12.483378 & 59696 & DBSP(P200)  \\
ZTF20aagwxlk & 153.42829 & 55.432205 & 59696 & DBSP(P200)\\
\textbf{ZTF20aagyaug} & 172.41282 & 36.883602 & 59603 & DBSP(P200) \\
ZTF20aakreaa & 181.52333 & 42.169888 & 59603 & DBSP(P200) \\
ZTF20aaorxzv & 132.39052 & 3.68048 & 59604 & LRIS(Keck I) \\
\textbf{ZTF20aaxwxgq} & 234.63610 & 46.126392 & 59696 & DBSP(P200)\\
\textbf{ZTF20abcvgpb} & 238.24977 & 21.046358 & 59696 & DBSP(P200) \\
ZTF20abgnlgv & 232.52715 & 7.172269 & 59696 & DBSP(P200) \\
\textbf{ZTF21aafkiyq} & 174.93478 & $-$1.727439 & 59603 & DBSP(P200) \\
\textbf{ZTF21abcsvbr} & 184.64829 & 18.771713 & 59603 & DBSP(P200) \\

    \hline
	\end{tabular}
	\end{threeparttable}
	\end{adjustbox}
\end{table}

\begin{figure*}
        \centering
            \begin{subfigure}{0.4\textwidth} 
            \includegraphics[width=\textwidth]{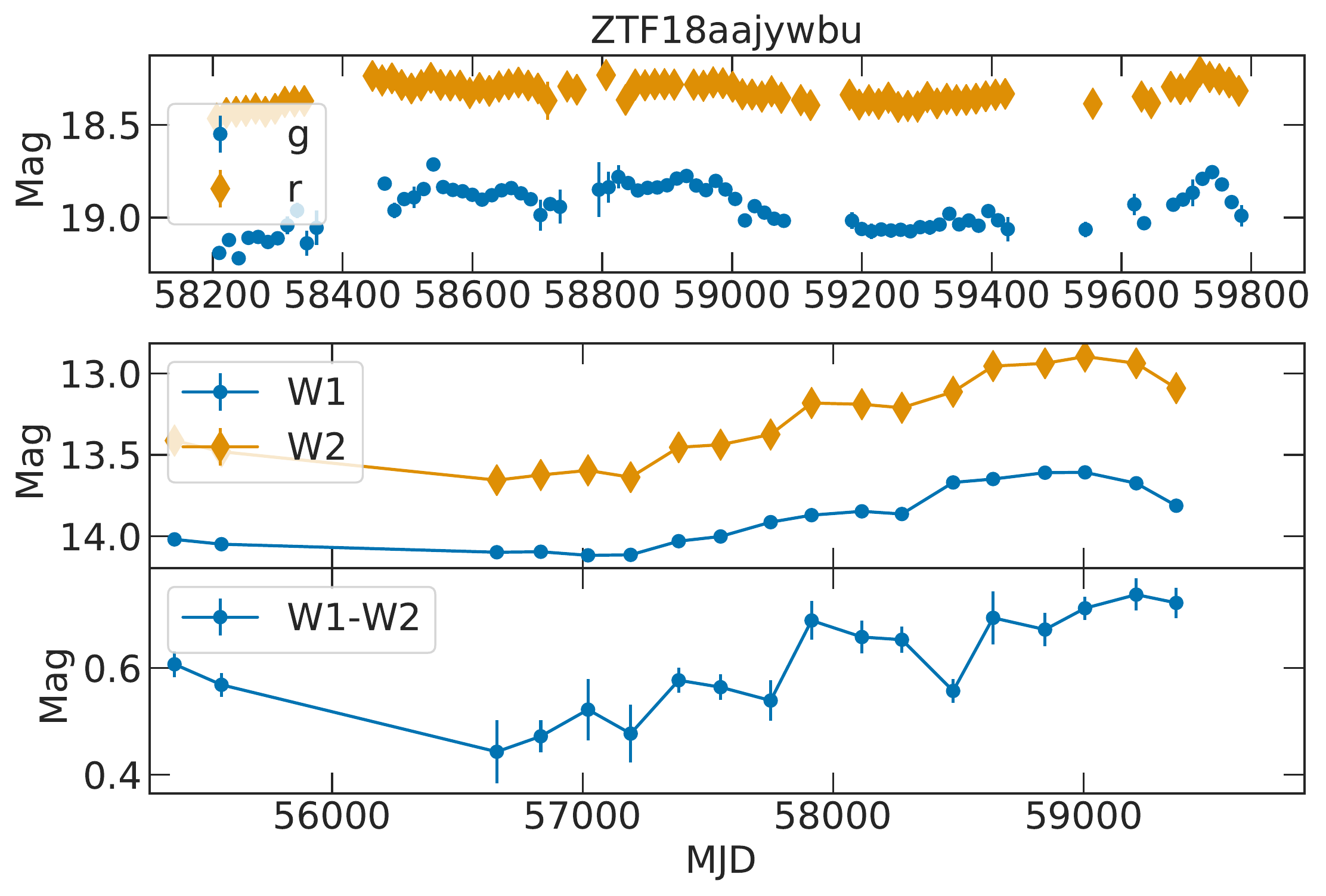}
        \end{subfigure}    
   \begin{subfigure}{0.4\textwidth} 
            \includegraphics[width=\textwidth]{plots/ZTF18aajywbu_spec.pdf}
        \end{subfigure}  

                    \begin{subfigure}{0.4\textwidth} 
            \includegraphics[width=\textwidth]{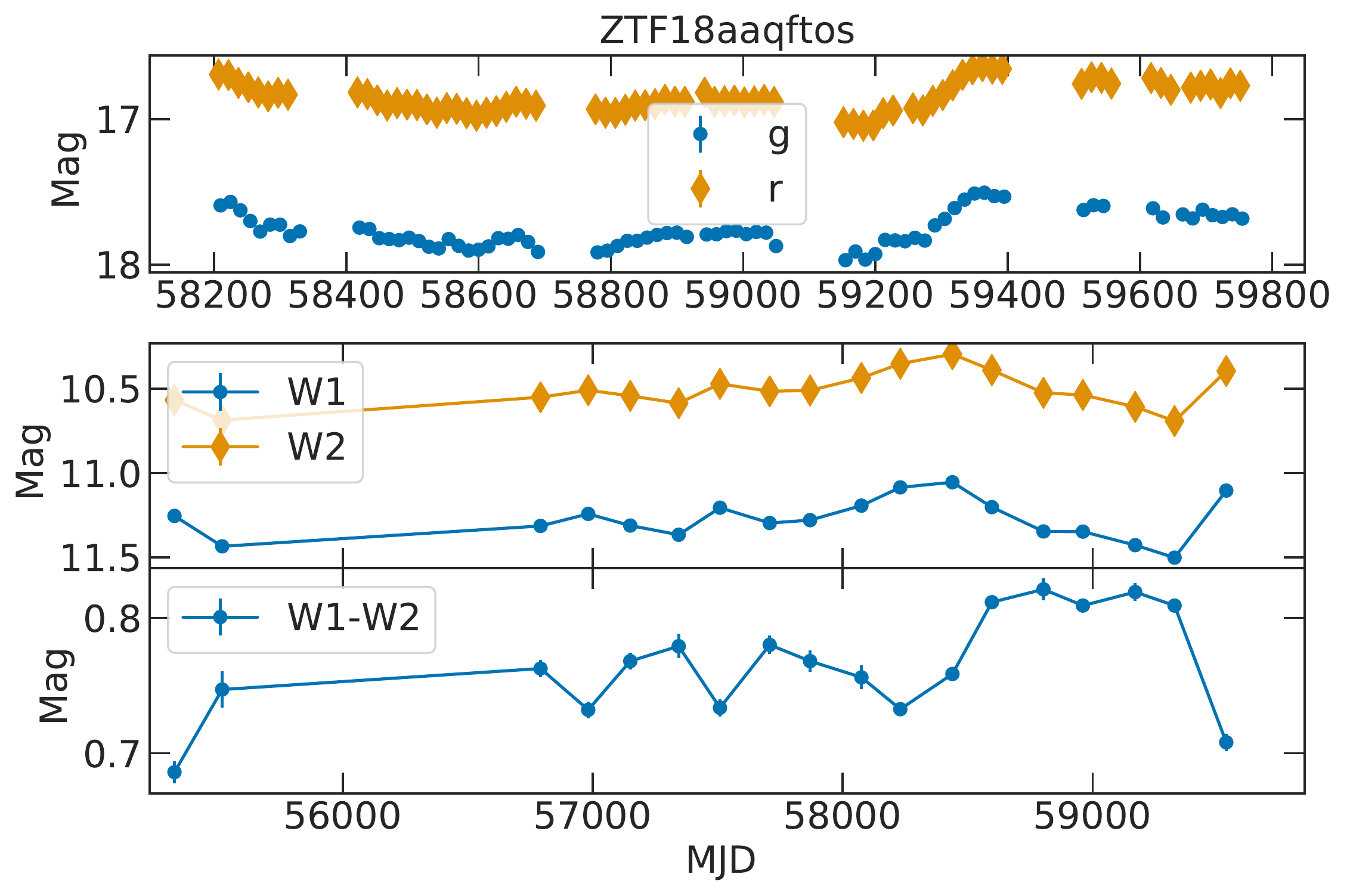}
        \end{subfigure}    
   \begin{subfigure}{0.4\textwidth} 
            \includegraphics[width=\textwidth]{plots/ZTF18aaqftos_spec.pdf}
        \end{subfigure} 

                    \begin{subfigure}{0.4\textwidth} 
            \includegraphics[width=\textwidth]{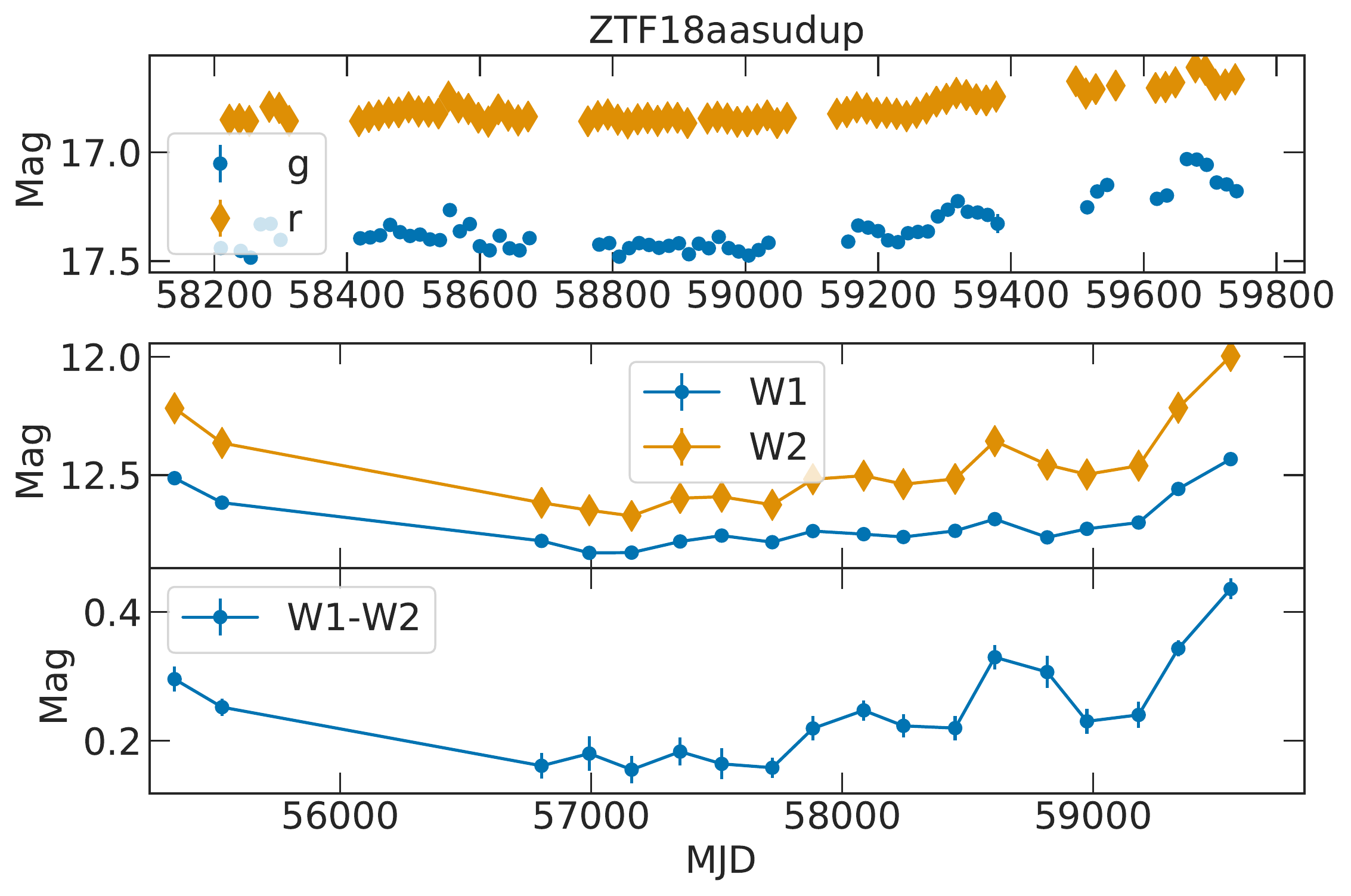}
        \end{subfigure}    
   \begin{subfigure}{0.4\textwidth} 
            \includegraphics[width=\textwidth]{plots/ZTF18aasudup_spec.pdf}
        \end{subfigure} 

                    \begin{subfigure}{0.4\textwidth} 
            \includegraphics[width=\textwidth]{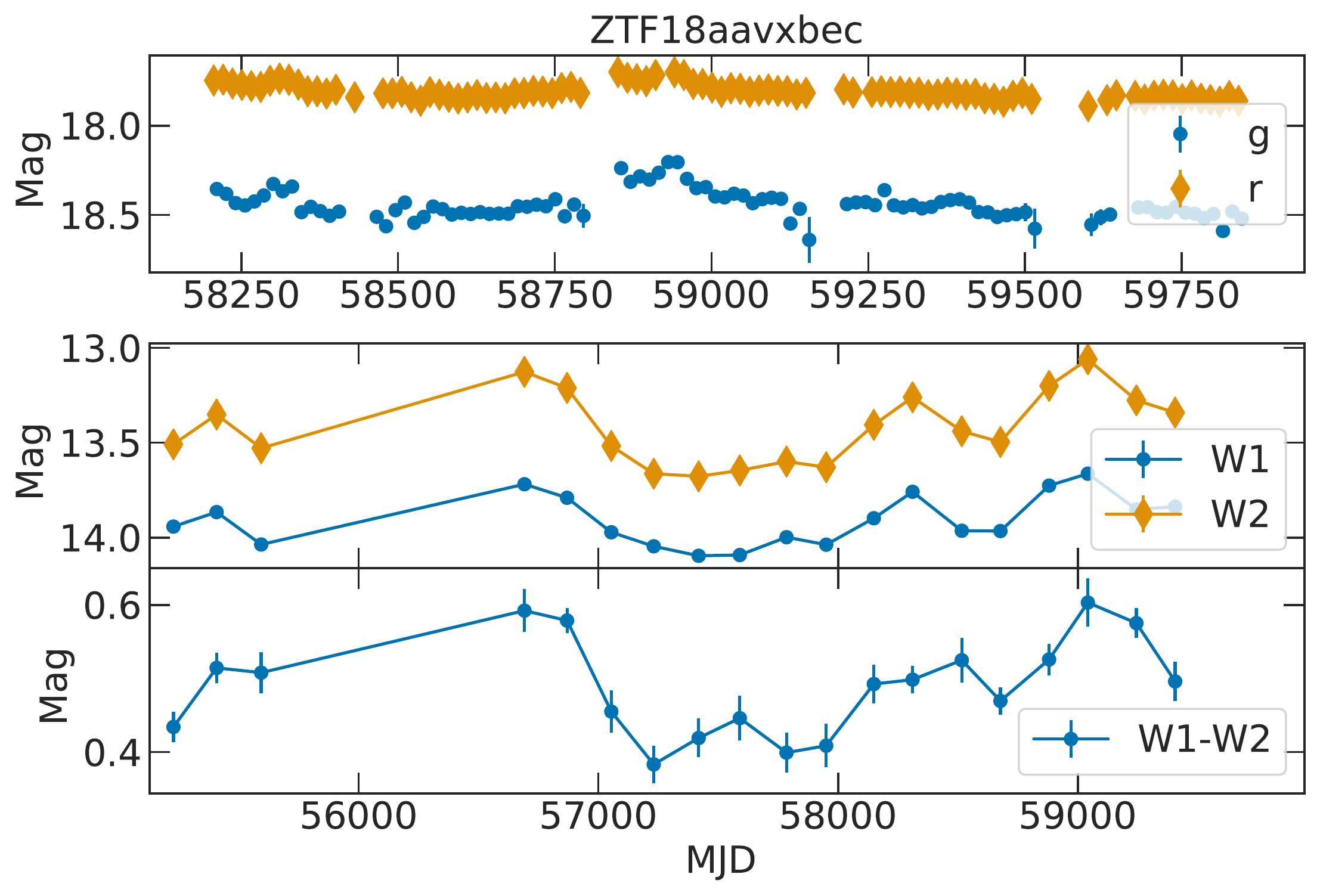}
        \end{subfigure}    
   \begin{subfigure}{0.4\textwidth} 
            \includegraphics[width=\textwidth]{plots/ZTF18aavxbec_spec.pdf}
        \end{subfigure} 

                    \begin{subfigure}{0.4\textwidth} 
            \includegraphics[width=\textwidth]{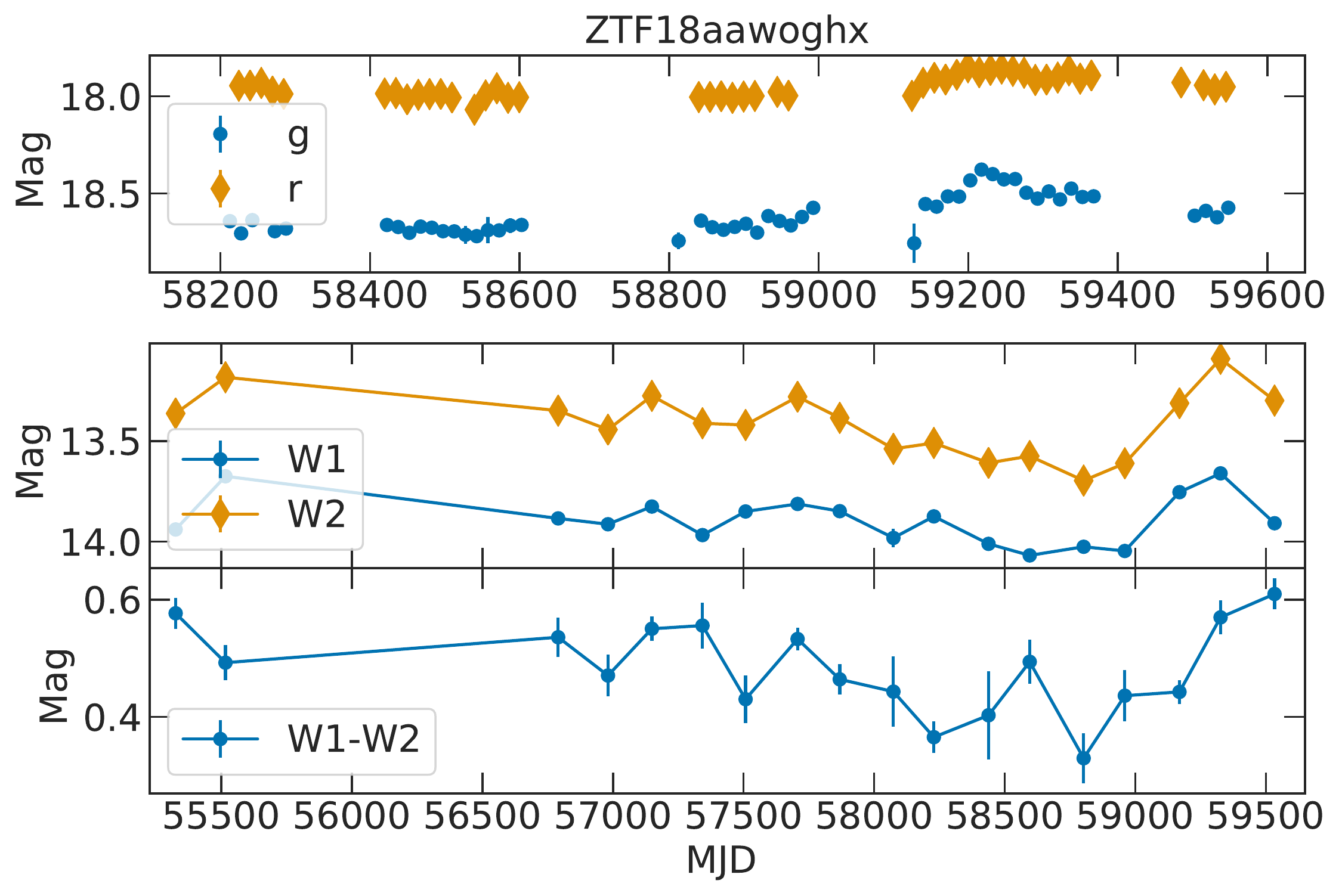}
        \end{subfigure}    
   \begin{subfigure}{0.4\textwidth} 
            \includegraphics[width=\textwidth]{plots/ZTF18aawoghx_spec.pdf}
        \end{subfigure}

\end{figure*}

\begin{figure*} \ContinuedFloat  
       \centering 
       \begin{subfigure}{0.4\textwidth} 
            \includegraphics[width=\textwidth]{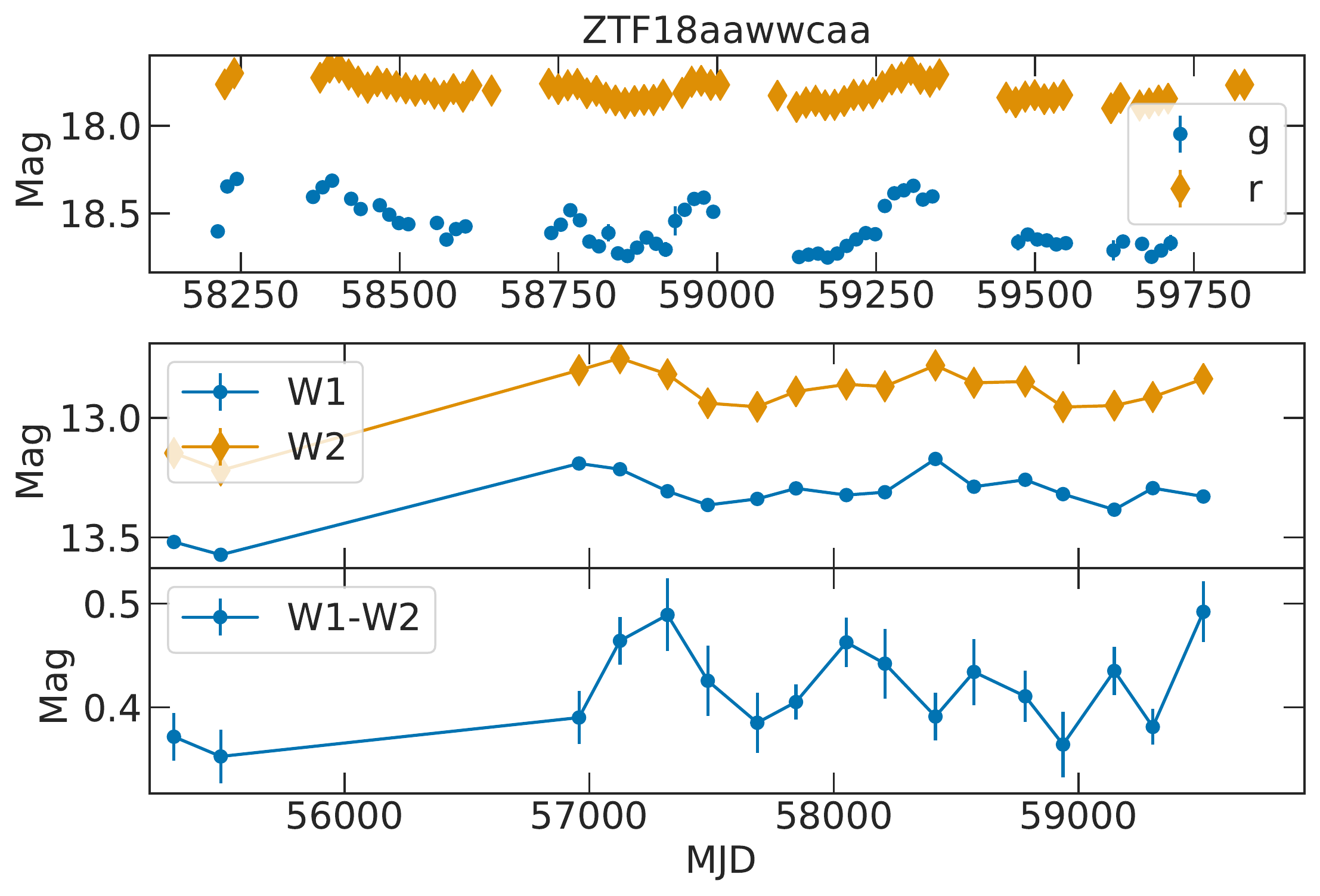}
        \end{subfigure}    
   \begin{subfigure}{0.4\textwidth} 
            \includegraphics[width=\textwidth]{plots/ZTF18aawwcaa_spec.pdf}
        \end{subfigure}

    \begin{subfigure}{0.4\textwidth} \ContinuedFloat
            \includegraphics[width=\textwidth]{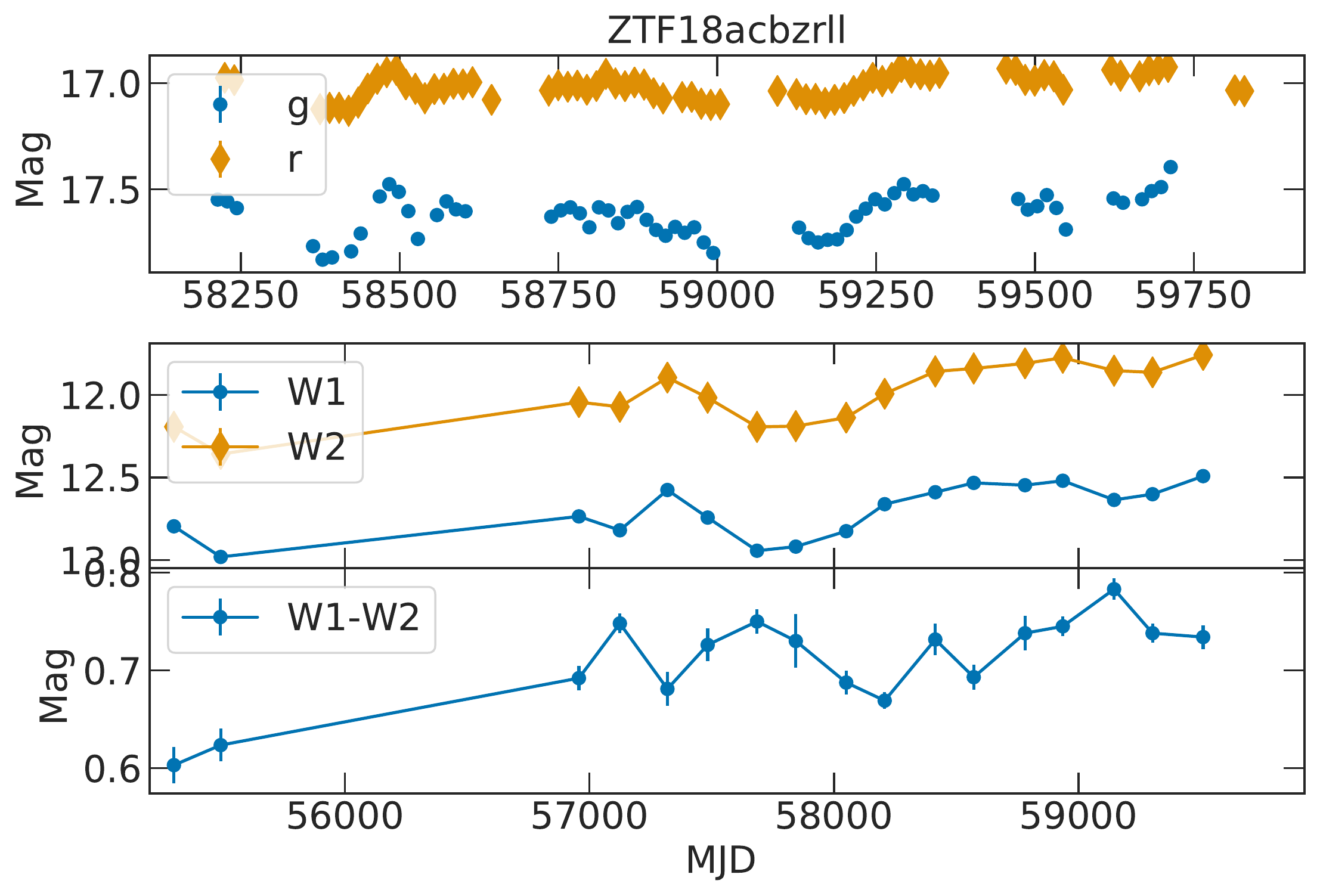}
        \end{subfigure}    
   \begin{subfigure}{0.4\textwidth} 
            \includegraphics[width=\textwidth]{plots/ZTF18acbzrll_spec.pdf}
        \end{subfigure} 

       \begin{subfigure}{0.4\textwidth} 
            \includegraphics[width=\textwidth]{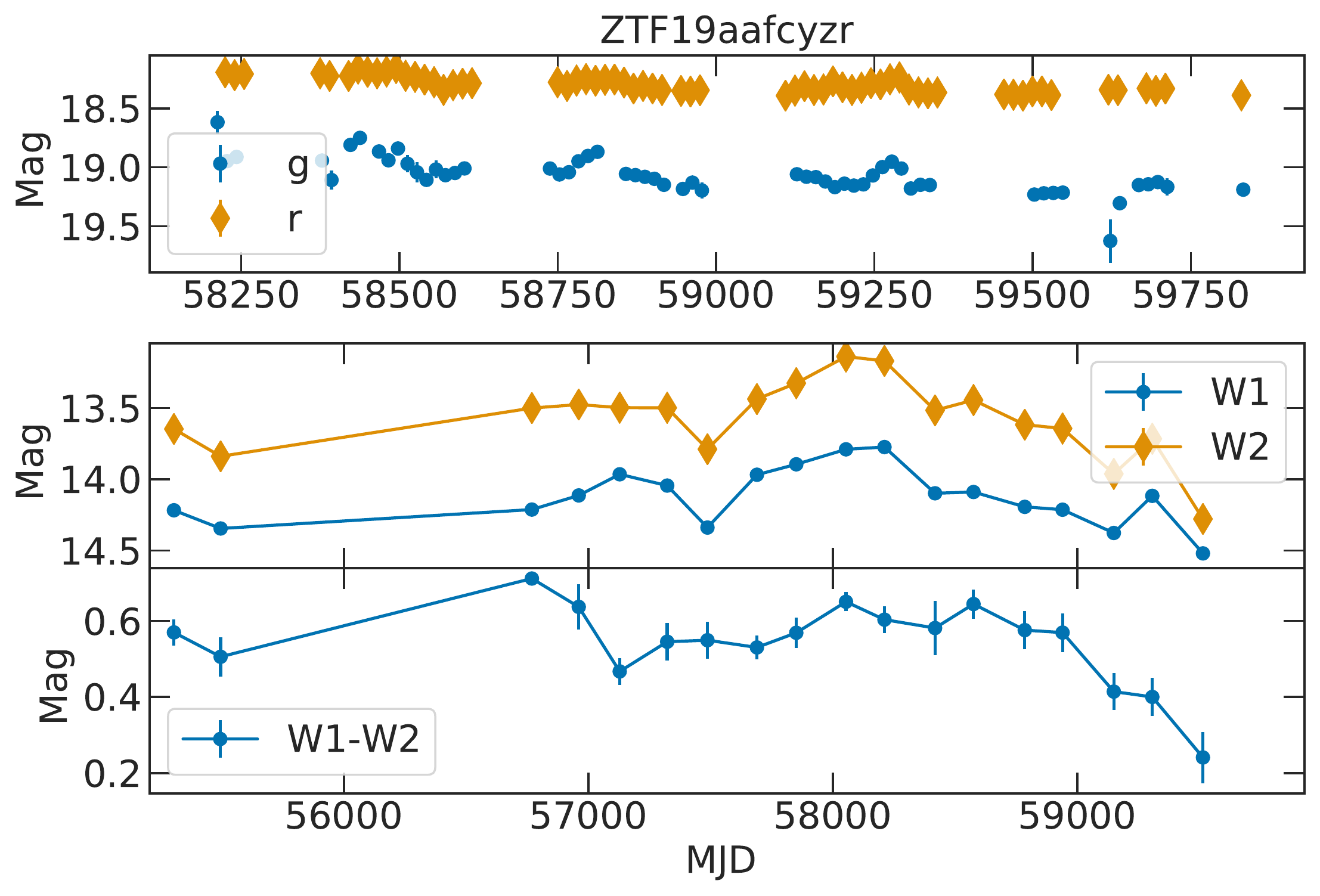}
        \end{subfigure}    
   \begin{subfigure}{0.4\textwidth} 
            \includegraphics[width=\textwidth]{plots/ZTF19aafcyzr_spec.pdf}
        \end{subfigure}

    \begin{subfigure}{0.4\textwidth} \ContinuedFloat
            \includegraphics[width=\textwidth]{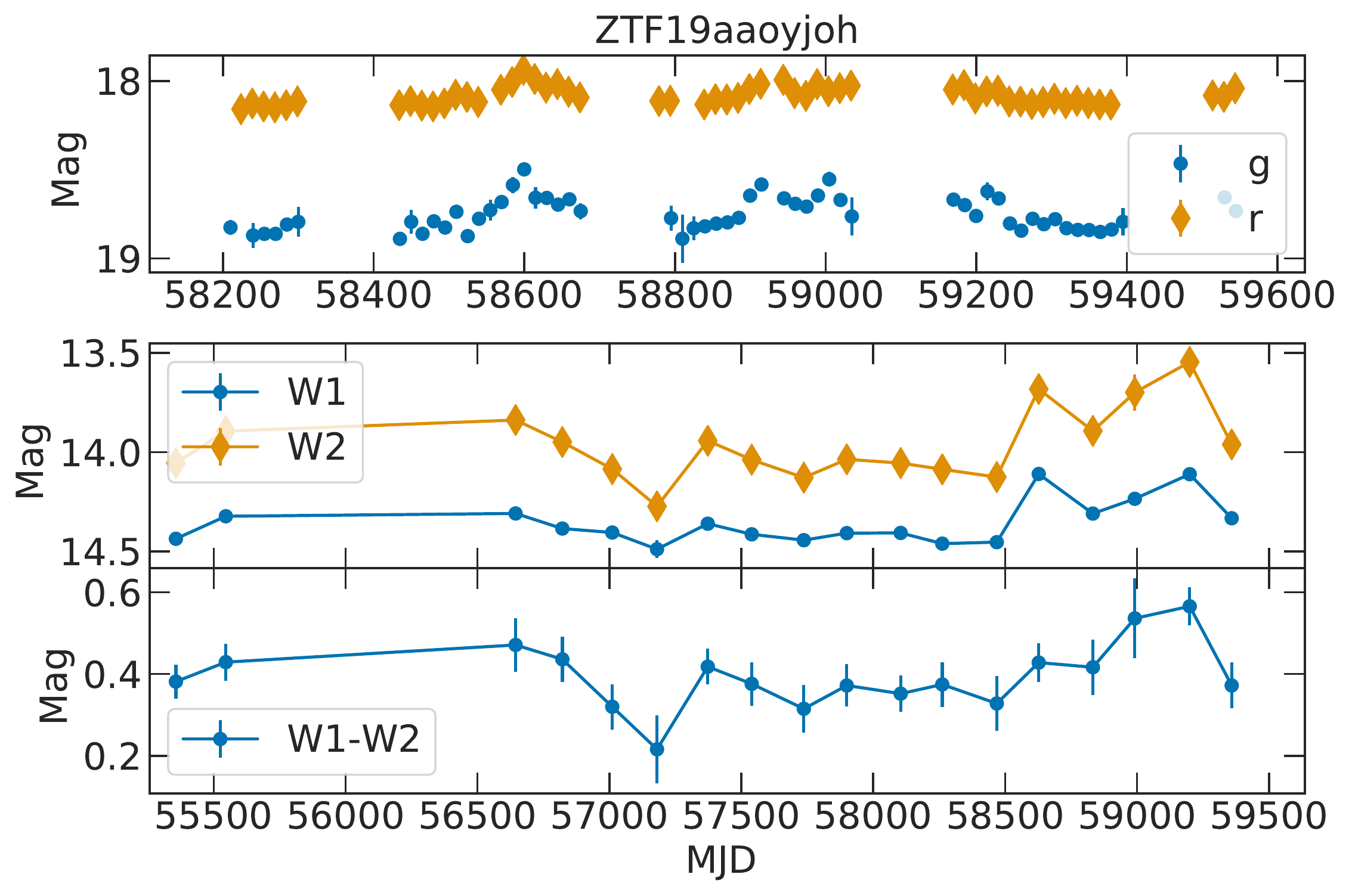}
        \end{subfigure}    
   \begin{subfigure}{0.4\textwidth} 
            \includegraphics[width=\textwidth]{plots/ZTF19aaoyjoh_spec.pdf}
        \end{subfigure} 

               \begin{subfigure}{0.4\textwidth} 
            \includegraphics[width=\textwidth]{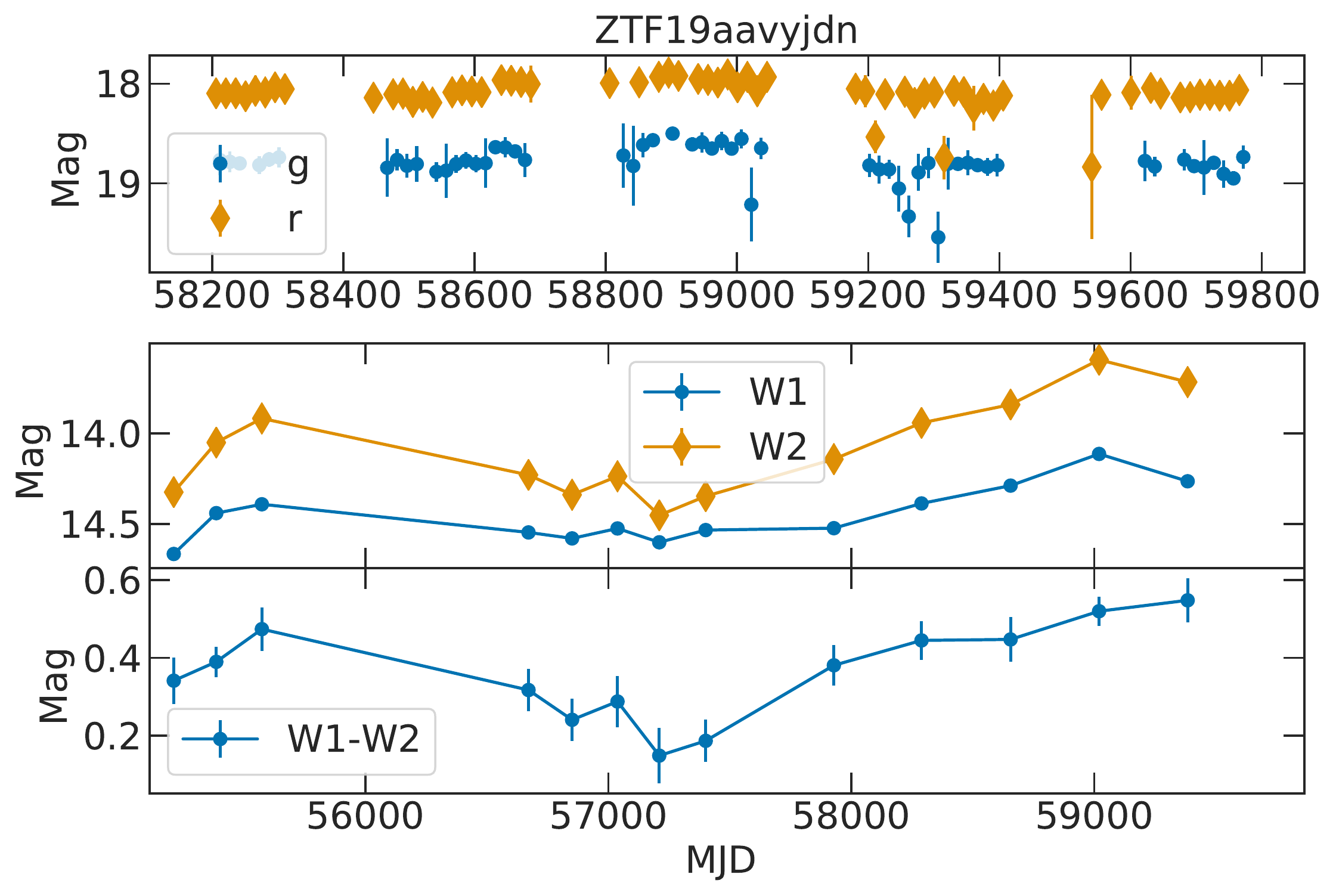}
        \end{subfigure}    
   \begin{subfigure}{0.4\textwidth} 
            \includegraphics[width=\textwidth]{plots/ZTF19aavyjdn_spec.pdf}
        \end{subfigure}

\end{figure*}

\begin{figure*} \ContinuedFloat  
       \centering

    \begin{subfigure}{0.4\textwidth} \ContinuedFloat
            \includegraphics[width=\textwidth]{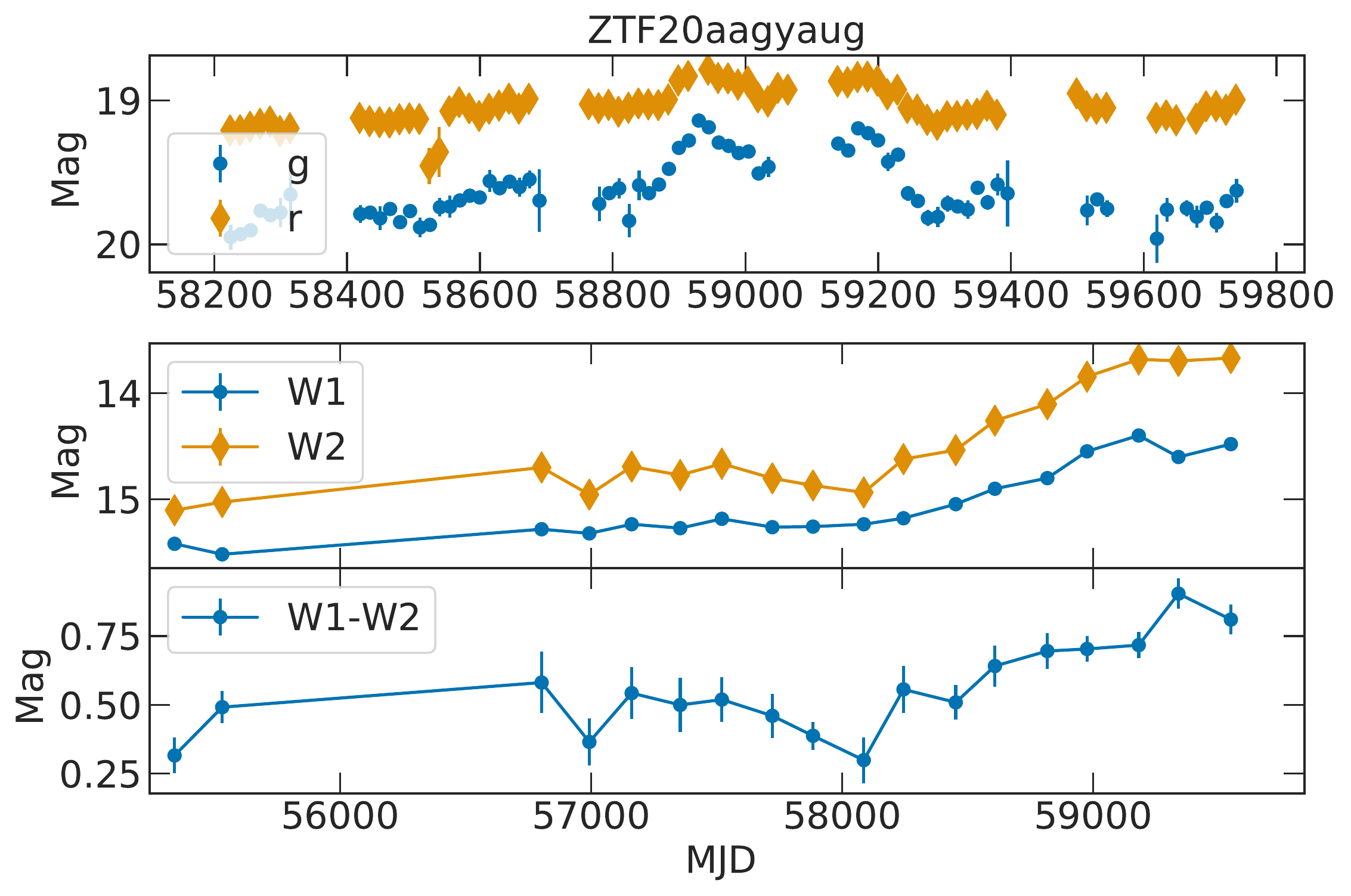}
        \end{subfigure}    
   \begin{subfigure}{0.4\textwidth} 
            \includegraphics[width=\textwidth]{plots/ZTF20aagyaug_spec.pdf}
        \end{subfigure} 

       \begin{subfigure}{0.4\textwidth} 
            \includegraphics[width=\textwidth]{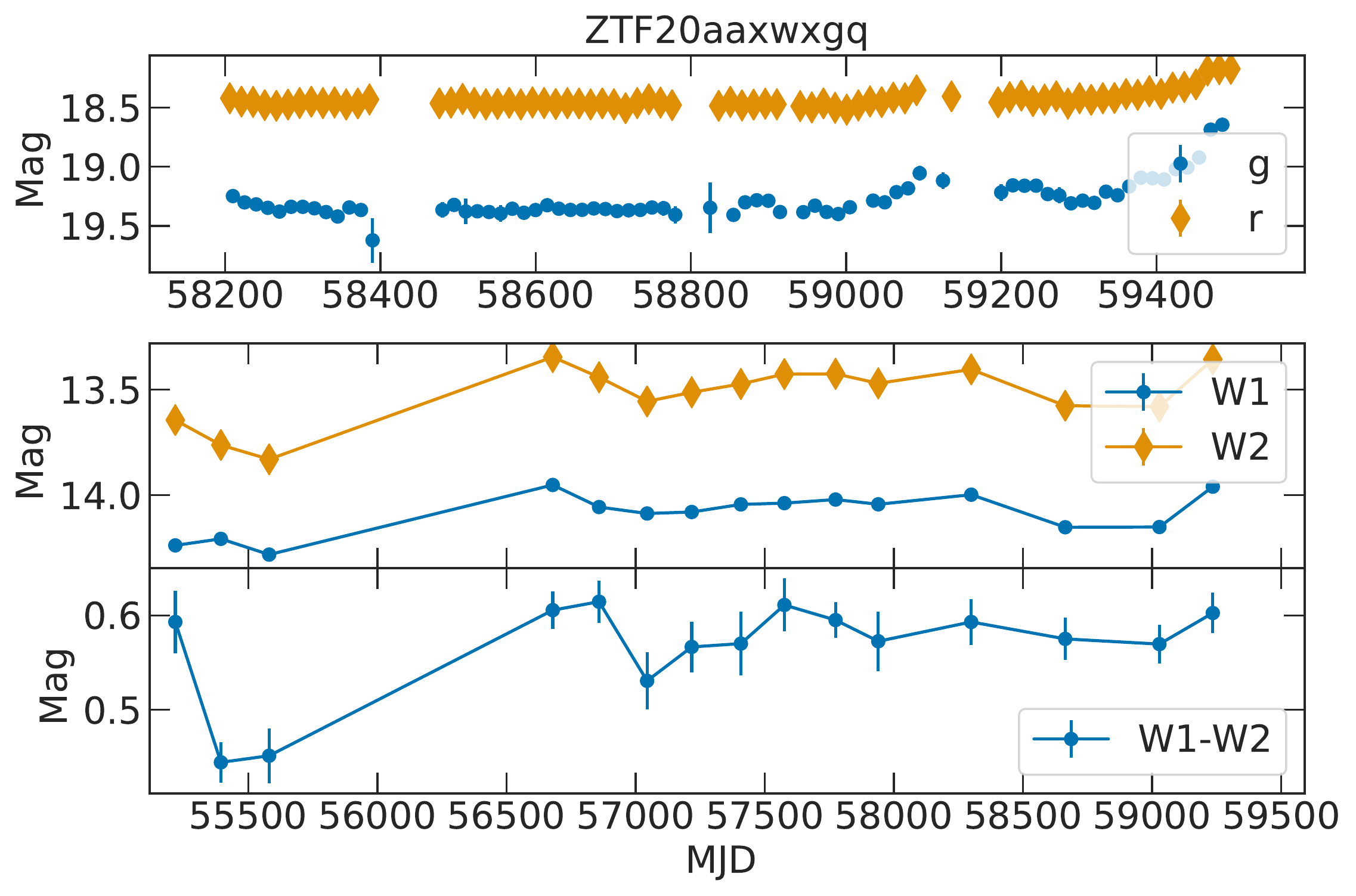}
        \end{subfigure}    
   \begin{subfigure}{0.4\textwidth} 
            \includegraphics[width=\textwidth]{plots/ZTF20aaxwxgq_spec.pdf}
        \end{subfigure}

    \begin{subfigure}{0.4\textwidth} \ContinuedFloat
            \includegraphics[width=\textwidth]{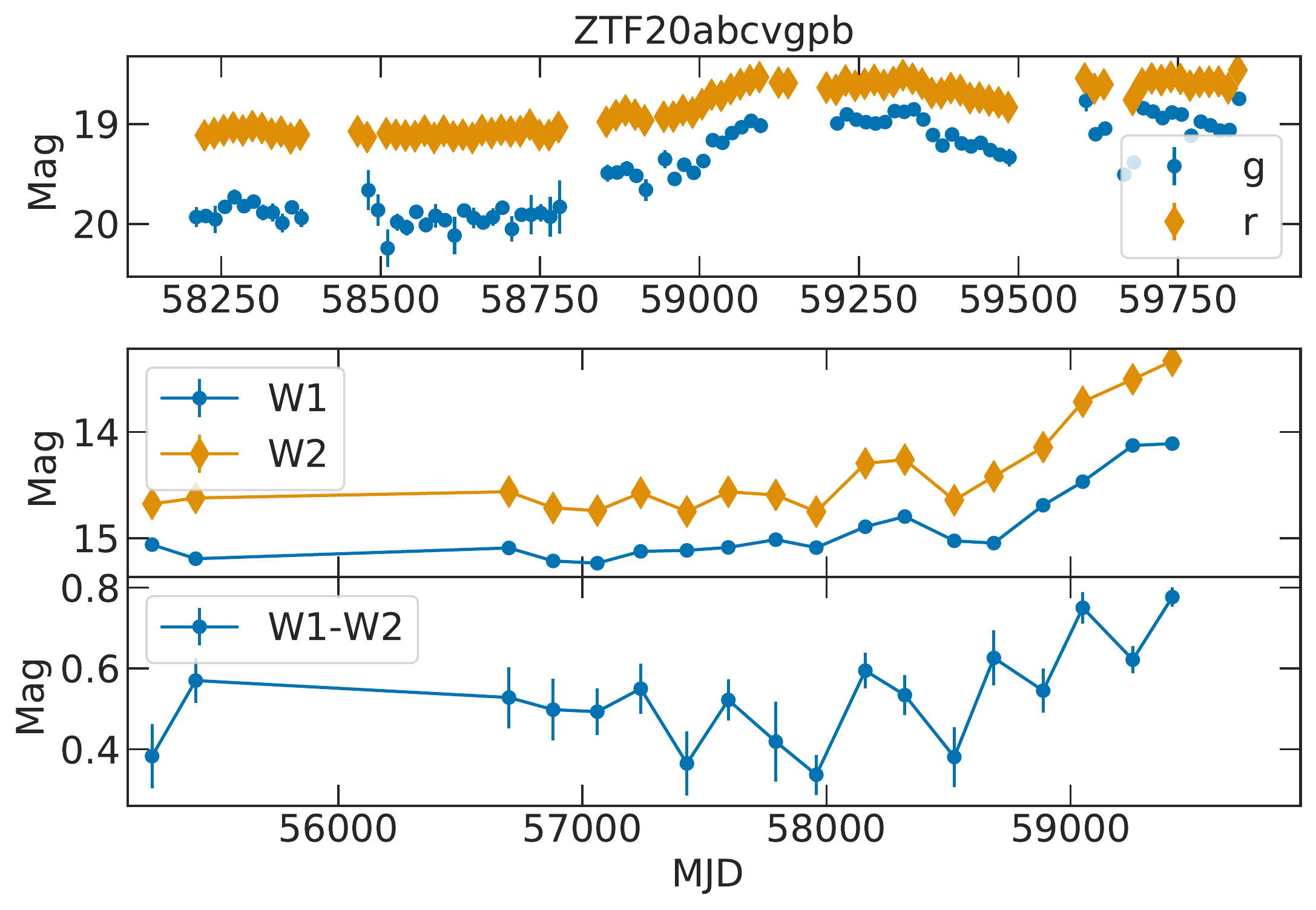}
        \end{subfigure}    
   \begin{subfigure}{0.4\textwidth} 
            \includegraphics[width=\textwidth]{plots/ZTF20abcvgpb_spec.pdf}
        \end{subfigure} 

               \begin{subfigure}{0.4\textwidth} 
            \includegraphics[width=\textwidth]{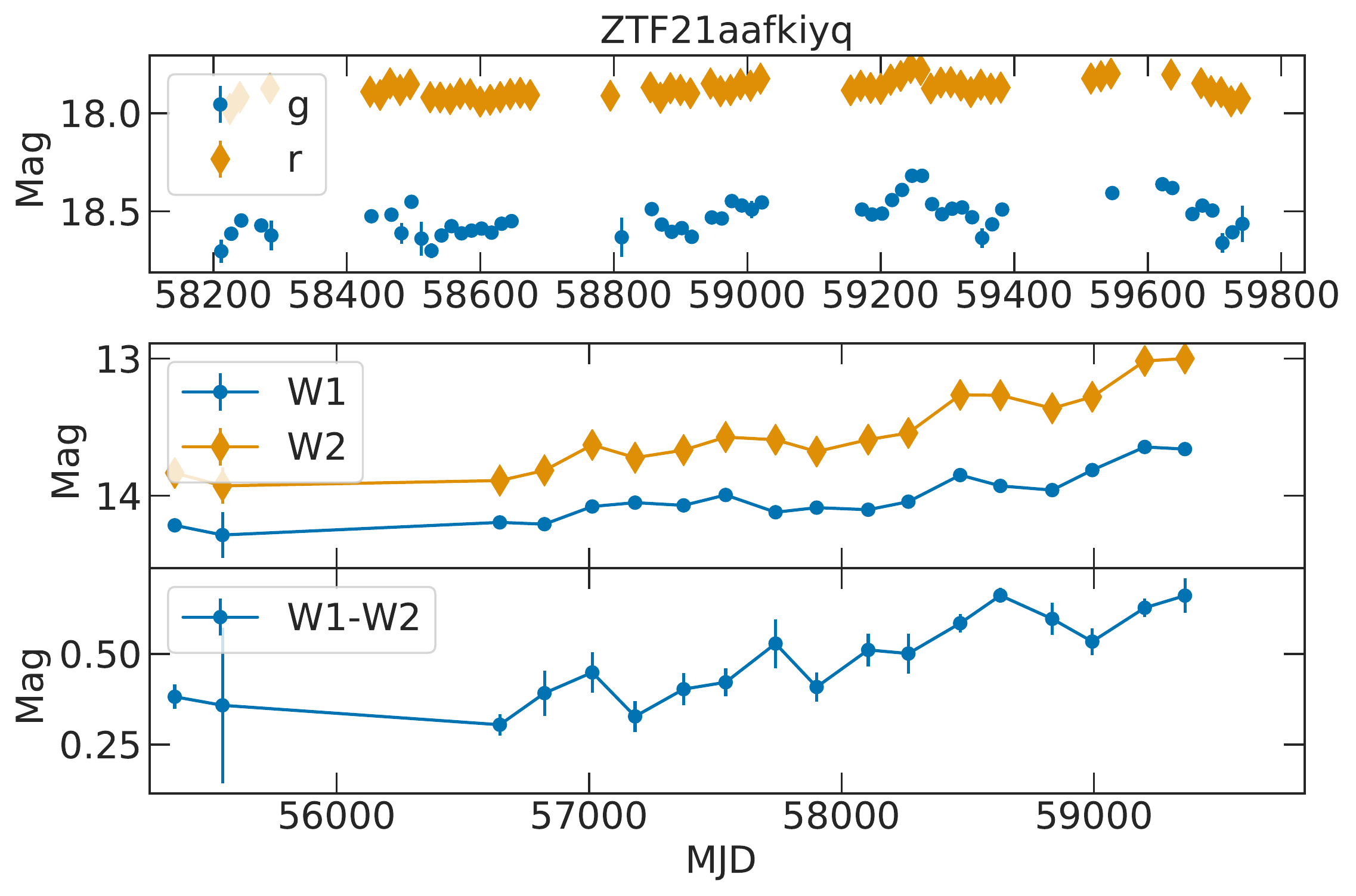}
        \end{subfigure}    
   \begin{subfigure}{0.4\textwidth} 
            \includegraphics[width=\textwidth]{plots/ZTF21aafkiyq_spec.pdf}
        \end{subfigure}   

\begin{subfigure}{0.4\textwidth} 
            \includegraphics[width=\textwidth]{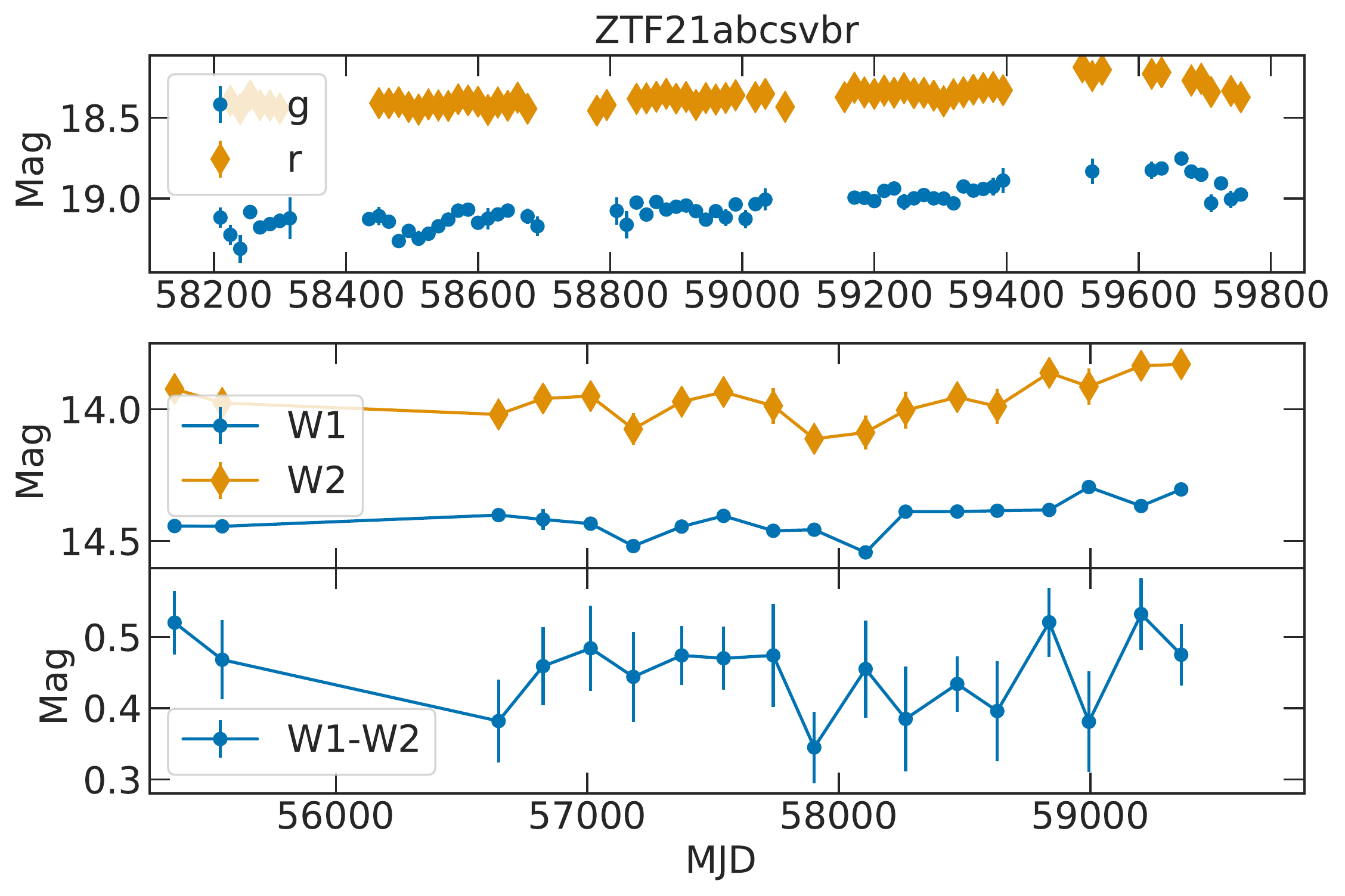}
        \end{subfigure}    
   \begin{subfigure}{0.4\textwidth} 
            \includegraphics[width=\textwidth]{plots/ZTF21abcsvbr_spec.pdf}
        \end{subfigure} 
     \caption{ZTF optical and \textit{WISE}-MIR light curves (left) and optical spectra (right) for the CL sources. } 
\label{fig: lc and spec}

\end{figure*}

\bsp	
\label{lastpage}
\end{document}